\documentclass{pasj02}
\usepackage{url}
\usepackage{bm}
\PassOptionsToPackage{pdftex}{graphicx}

\Received{$\langle$reception date$\rangle$}
\Accepted{$\langle$acception date$\rangle$}
\Published{$\langle$publication date$\rangle$}

\begin{document}

\title{Wind-driven angular momentum removal and X-ray ionization effects in Roche-lobe overflowing HMXBs}

\author{Atsuo T. Okazaki$^1$ and Stanley P. Owocki$^2$}

\altaffiltext{1}{Faculty of Engineering, Hokkai-Gakuen University, Sapporo 062-0911, Japan}
\altaffiltext{2}{Department of Physics \& Astronomy, University of Delaware, Newark, DE 19716, USA}
\email{okazaki@hgu.jp}


\maketitle

\begin{abstract}
We investigate the dynamical interaction between stellar winds and Roche-lobe overflow (RLO) streams in high-mass X-ray binaries using three-dimensional SPH simulations.
We show that the donor wind can exert a strong dynamical influence on the accretion flow. In the absence of X-ray ionization, the wind interacts asymmetrically with the disk and carries retrograde angular momentum, resulting in a net removal of angular momentum and a reduction of the disk size.
When X-ray photoionization is included, the response of the system becomes strongly non-monotonic. At high luminosity ($L_\mathrm{X} = 10^{38}\;\mathrm{erg\ s}^{-1}$), wind acceleration is suppressed and the flow approaches the no-wind case. In contrast, at moderate luminosity ($L_\mathrm{X} = 10^{37}\;\mathrm{erg\ s}^{-1}$), the system enters a qualitatively different accretion regime.
In this regime, the partially ionized wind becomes dense and dynamically important, strongly perturbing the RLO stream. As a result, the accretion flow transitions from RLO-dominated to wind-dominated, and the RLO stream undergoes a dynamical bifurcation into low- and high-angular-momentum branches. This leads to suppressed disk formation despite the presence of a strong mass supply.
These results demonstrate that moderate ionization can be more disruptive than both weak and strong ionization, providing a new mechanism regulating accretion flows in RLO HMXBs.
\end{abstract}


\section{Introduction}
\label{sec:intro}

High-mass X-ray binaries (HMXBs) consist of a compact object 
(neutron star or black hole) accreting matter from a massive 
OB-type companion star. 
Depending on the dominant mass-transfer mechanism, 
HMXBs are usually classified into three main categories:
Be X-ray binaries, supergiant X-ray binaries, and 
Roche-lobe overflowing systems.

Be X-ray binaries constitute roughly half of the known HMXBs. 
In these systems the compact object accretes matter from the 
circumstellar decretion disk of a rapidly rotating Be star 
(e.g., \cite{Reig2011}). 
Supergiant X-ray binaries account for about 40\% of the population 
and are typically wind-fed systems in which the compact object 
captures part of the strong radiatively driven wind from an 
OB supergiant (e.g., \cite{Martinez-Nunez2017}).

A third class consists of Roche-lobe overflowing (RLO) HMXBs, 
in which mass transfer occurs through the inner Lagrange point 
from the donor star to the compact object. 
These systems are relatively rare but include several 
well-studied objects such as Cen X-3, SMC X-1, and LMC X-4.
They occupy a distinct region in the Corbet diagram 
(\cite{Corbet1986}), characterized by short spin periods 
and relatively short orbital periods.

In low-mass X-ray binaries (LMXBs), Roche-lobe overflow typically 
produces a stable accretion disk around the compact object. 
However, the situation in HMXBs could be fundamentally different. 
Because the donor stars are massive OB stars, they drive strong 
line-driven winds with mass-loss rates of 
$\sim10^{-7}$--$10^{-6} M_\odot\;\mathrm{yr}^{-1}$ 
(e.g., \cite{Vink2001}). 
Therefore, Roche-lobe overflow in HMXBs could occur in a 
strong-wind environment.

The interaction between stellar winds and accretion disks 
has been studied in several contexts, including wind-fed 
accretion flows (e.g., \cite{Blondin1990}) and circumstellar disk ablation 
by stellar winds \citep{Kee2016}. 
Recent studies have investigated Roche-lobe overflow in high-mass X-ray binaries, 
including models that incorporate stellar winds (e.g., \cite{Dickson2024}). 
In these studies, however, the wind is not treated as a primary dynamical agent 
interacting with the RLO stream. 
In particular, the role of the wind in angular momentum transport and its 
direct impact on the structure of the accretion flow remain largely unexplored.
This motivates a study in which the stellar wind is treated as an active 
dynamical component that can modify the RLO flow through direct interaction.

Another important physical ingredient in RLO HMXBs is their strong X-ray emission.
X-ray irradiation from the compact object can significantly modify the
donor wind through photoionization (e.g., \cite{Hatchett-McCray1977}). By reducing line-driving efficiency,
X-ray ionization is expected to suppress wind acceleration, thereby
altering both the mass-loss rate and the velocity structure of the wind (e.g., \cite{Stevens-Kallman1990}).

Previous studies of X-ray irradiated winds in HMXBs have mainly addressed the ionization structure of the donor wind, wind inhibition, and wind-fed accretion onto the compact object (e.g., \cite{Hatchett-McCray1977, Blondin1990, Stevens-Kallman1990}). In RLO HMXBs, however, a dense RLO stream and a disk-like bound flow coexist with the donor wind. The central question of this work is therefore how the donor wind, modified by X-ray photoionization, interacts dynamically with the RLO stream and changes the angular momentum budget, disk formation, and accretion regime.

Using three-dimensional SPH simulations, we investigate in this work how the stellar wind and X-ray photoionization affect accretion flows in RLO HMXBs.
We consider four representative cases: no ionization with and without wind, moderate ionization
($L_\mathrm{X} = 10^{37}\;\mathrm{erg\ s}^{-1}$), and strong ionization
($L_\mathrm{X} = 10^{38}\;\mathrm{erg\ s}^{-1}$). 
While the RLO mass-transfer rate is kept fixed among all models, 
the X-ray luminosity $L_\mathrm{X}$ is treated as an externally specified parameter. 
This allows us to isolate the dynamical effect of X-ray ionization on the donor wind 
and its interaction with the RLO stream.

This paper is organized as follows. In Section~2 we describe the numerical
setup and the implementation of X-ray ionization. Section~3 presents the
global flow structures for the above four cases, and then
analyzes wind torques on accretion flows and the ionization effect on the wind-disk dynamics.
In Section~4 we discuss the physical implications, and in
Section~5 we summarize our conclusions.

\section{Numerical Method}
\label{sec:method}

\begin{figure}
\begin{center}
\includegraphics[width=0.8\hsize]{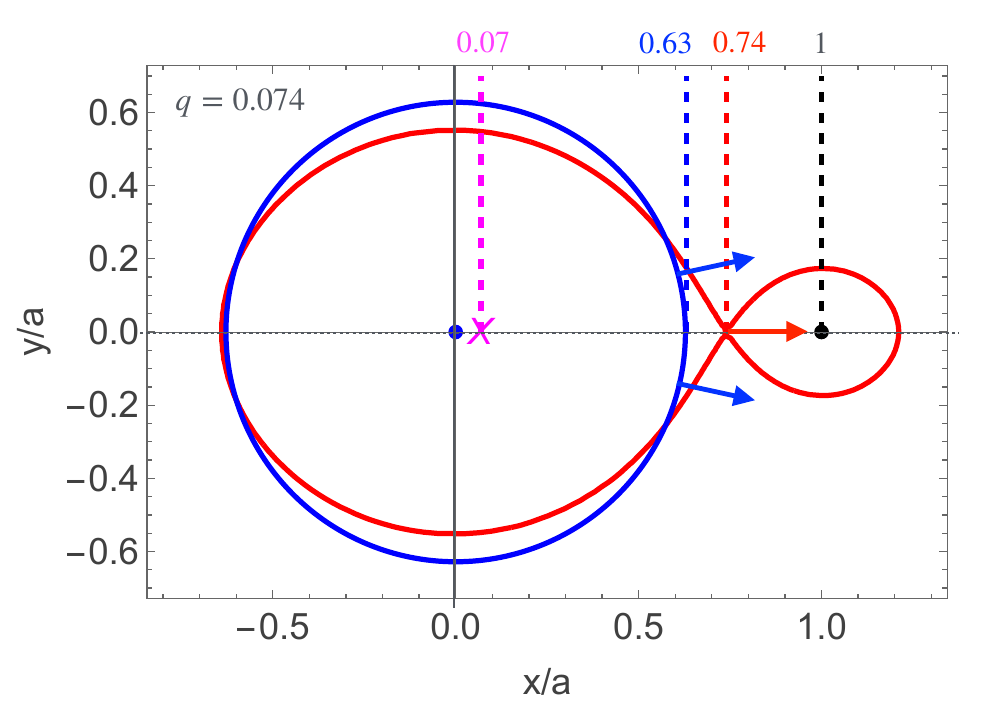}
\end{center}
\caption{
The Roche geometry and the adopted wind and RLO injection geometry. The origin of the coordinates is set at the center of the donor star. The binary mass ratio $q=M_\mathrm{NS}/M_{*}$ is annotated in the upper-left corner. In units of the semi-major axis $a$, the adopted spherical donor radius is $R_{*}/a = 0.63$, slightly larger than the volume-equivalent Roche-lobe radius, while the L$_1$ point is located at $x/a = 0.74$. The red cross marks the center of mass of the binary. The blue and red arrows indicate the wind-launching radius and the RLO injection position, respectively.
{Alt text: Schematic view of the adopted binary geometry in the orbital plane. The donor star is centered near the origin, the neutron star is located to the right, and the Roche geometry, center of mass, wind-launching position, and RLO injection direction are indicated. The figure illustrates that the adopted donor radius nearly fills the Roche lobe and that mass transfer occurs near the L1 point.}}
\label{fig:binary_geometry}
\end{figure}

We consider a binary comprising an O star and a neutron star in a circular orbit. Because of the close orbit, the rotation of the O star is synchronous with that of the orbital motion. 
For binary parameters, we adopt values representative of the RLO HMXB Cen X-3. 
The donor star has mass $M_{*} = 20 M_\odot$ \citep{vanderMeer2007} and radius $R_{*} = 12 R_\odot$ \citep{Naik2011}, with an effective temperature 
$T_\mathrm{eff} = 39,000\,\mathrm{K}$ appropriate for an O6--7 giant 
\citep{Hutchings1979, Martins2005}. 
The neutron star mass and radius are taken to be 
$M_\mathrm{NS} = 1.49 M_\odot$ \citep{vanderMeer2007} and $R_\mathrm{NS} = 10^6\;\mathrm{cm}$, respectively. The orbital period, $P_\mathrm{orb}$, is $P_\mathrm{orb} = 2.09\;\mathrm{d}$ \citep{Klawin2023}, and the semi-major axis, $a$, is $a = 1.33 \cdot 10^{12}\;\mathrm{cm}$.
In this study, we assume that the donor rotates at the co-rotation frequency of $\Omega_{*} = \Omega_\mathrm{orb} = 0.516 \Omega_\mathrm{crit}$, where $\Omega_\mathrm{orb} = [G (M_{*} + M_\mathrm{NS})/a^3]^{1/2}$ and $\Omega_\mathrm{crit} = (GM_{*}/R_{*}^3)^{1/2}$ are the orbital frequency of the binary and the spherical critical angular frequency of the donor, respectively.
The Roche geometry and the adopted wind and RLO injection geometry are shown in Figure~\ref{fig:binary_geometry}.

\subsection{SPH setup}
\label{sec:sph}

We model the interaction between the stellar wind and the Roche lobe overflow by performing three-dimensional Smoothed Particle Hydrodynamics (SPH) simulations. The code is based on a version originally developed by M.\ R.\ Bate and his colleagues \citep{Bate1995}, and is basically identical to that used by \citet{OkazakiRussell2014}, except that the current version has a Roche lobe overflow (RLO) to study RLO HMXBs.

In our code, the wind and the Roche lobe overflow are modeled by ensembles of gas particles with different particle masses, whereas the binary star components are modeled by two sink particles with appropriate gravitational masses and specified accretion radii. We set the accretion radius of the O star to $R_{*}$, but for the neutron star we adopt an accretion radius of $10^{-3} a$ ($\sim 10^{9}\;\mathrm{cm} \gg R_\mathrm{NS}$), because of our limited spatial resolution. Gas particles which fall within one of these accretion radii are accreted by the sink particle.
The binary is assumed to orbit in the $xy$-plane.

In the simulations shown in this paper, the artificial viscosity parameter $\alpha_\mathrm{SPH}$ is variable with time and space so as to emulate the Shakura-Sunyaev (\yearcite{Shakura-Sunyaev1973})'s viscosity parameter $\alpha_\mathrm{SS} = 0.1$ \citep{Okazaki2002, Rubio2025}. The other artificial viscosity parameter $\beta_\mathrm{SPH}$ is set to 0. 
This prescription is applied to both RLO-origin and wind-origin particles. Therefore, in disk-like regions it provides an effective Shakura-Sunyaev viscosity with $\alpha_\mathrm{SS} = 0.1$, whereas in wind-stream interaction regions it should be regarded primarily as an adaptive shock-capturing/numerical-dissipation prescription rather than as a physical shear viscosity.
As for cooling, the code takes into account radiative cooling with the cooling function generated by CLOUDY 90.01 for an optically thin plasma with solar abundances \citep{Ferland1996}. The floor temperature is set to the stellar effective temperature. Although our floor temperature is rather high, it would have little effect on the dynamical nature of the wind-disk interaction, the focus of this paper.

\subsection{Stellar wind model}
\label{sec:wind_model}

In order to model radiatively accelerating winds, we have introduced an external force into the code to emulate the $\beta$-velocity law, $v_\mathrm{w}(r) = v_{\infty} (1-R_{*}/r)^{\beta}$, where $\beta=1$ is adopted and the terminal velocity $v_{\infty}$ is set to $2.6 v_\mathrm{esc}$ \citep{Vink2001}, with $v_\mathrm{esc}=(2GM_{*}/R_{*})^{1/2}$ being the escape velocity of the donor. 
We inject wind particles at a radius where the wind is already supersonic at $v_\mathrm{w} = 2 c_\mathrm{s}$, where $c_\mathrm{s}$ is the sound speed. The initial wind temperature is set to $T_\mathrm{eff}$.
The wind mass-loss rate, which is kept constant, is estimated by the mass-loss recipe of \citet{Bjorklund2023} to be $\dot{M}_\mathrm{w} \sim 4.2 \times 10^{-7} M_{\odot}\,\mathrm{yr^{-1}}$.
In order to optimize the resolution and computational efficiency, the wind particles are launched only in a narrow range of azimuthal and vertical angles toward the neutron star, i.e., azimuthally from $-60$ degrees to $+60$ degrees with respect to the direction of the neutron star, and vertically from 45 degrees below the orbital plane to 45 degrees above it. This restriction is adopted to focus computational resolution on the region relevant to wind-stream interaction.

In addition, in order to avoid direct overlap between the injected wind and the RLO stream near the L$_1$ point, we exclude a narrow angular region around the L$_1$ direction from wind injection. The half-opening angle of this no-wind region is set to 3 degrees, which is larger than the RLO injection cone ($\theta_\mathrm{rlof} = 1^\circ$), ensuring that the RLO stream is not artificially disrupted at injection.
In actual RLO systems, the effective gravity is reduced near the substellar region facing the compact object (e.g., \cite{Espinosa-Rieutord2012}). By analogy with gravity darkening, this tidal reduction of the effective gravity may reduce the local wind mass flux near the L$_1$ direction. Although this effect is not modeled self-consistently in the present simulations, it provides a physical motivation for excluding a narrow no-wind cone around the RLO stream.

Since the donor star rotates at the co-rotation frequency $\Omega_{*}$, the wind particles are launched with the latitude-dependent, tangential velocity of $R_{*} \Omega_{*} \sin \theta$, where $\theta$ is the polar angle. Also, to make the non-ionized wind model consistent with the gravity darkening effect (e.g., \cite{vonZeipel1924}), the initial wind density is adapted for the local wind mass-loss rate to vary as $\dot{m}_\mathrm{w}(\theta) \propto 1-(\Omega_{*}/\Omega_\mathrm{crit})^2 \sin^2 \theta$ \citep{Owocki2004}, where $\Omega_{*}/\Omega_\mathrm{crit} = 0.516$ is adopted, as mentioned above.

\subsection{Roche-lobe overflow mass injection}

Mass transfer via Roche-lobe overflow (RLO) is implemented by injecting SPH particles in the vicinity of the inner Lagrange point (L1). Because the detailed hydrodynamics near L$_1$ is not resolved in our simulations, the injection must be prescribed in an approximate manner. 
In the main text, we focus on the cone-injection model as our fiducial
RLO prescription.\footnote{We tested alternative injection prescriptions and confirmed that 
the qualitative results are unchanged.}


In the fiducial model, particles are injected from a small region located slightly outside the L$_1$ point, within a conical volume oriented toward the neutron star. The cone is characterized by a half-opening angle $\theta_\mathrm{rlof}$ of 1 degree, and particles are launched with an initial radial velocity $v_\mathrm{rlof} = 2 c_\mathrm{s}$, defined in the co-rotating frame of the binary system. 
The injection region is centered just outside of the L$_1$ point and co-rotates with the binary.

This prescription is intended to represent a moderately collimated stream emerging from the vicinity of L$_1$ while allowing for a finite spread in direction. The cone geometry provides a simple and flexible way to control the angular width of the injected stream without introducing additional assumptions about the unresolved microphysics at L$_1$.

The mass injection rate is set to reproduce desired X-ray luminosities ($10^{38}\;\mathrm{erg\ s}^{-1}$) in the steady state limit if it is converted to the same rate of mass accretion. Particles are inserted continuously in time to ensure a quasi-steady supply of material. Unless otherwise stated, all results presented in this paper are based on this cone injection model.

\subsection{Photoionization model}
\label{sec:ion_model}

X-ray photoionization reduces the efficiency of line driving by 
removing line opacity \citep{Stevens-Kallman1990}, 
thereby suppressing wind acceleration in HMXBs \citep{Blondin1990}.
We adopt a simple parametric form to mimic this effect
by reducing the wind acceleration by the following factor
\begin{equation}
   f_\mathrm{ion} = \frac{1}{1 + \left( \frac{\xi}{\xi_0} \right)^{\alpha_{\xi}}},
\label{eq:fion}
\end{equation}
where $\xi$ is the ionization parameter defined by
\begin{equation}
   \xi = \frac{L_\mathrm{X}}{n r^2}
\label{eq:xi}
\end{equation}
with $L_\mathrm{X}$ and $n$ being the X-ray luminosity and the local number density, respectively \citep{Tarter1969}.
We adopt $\xi_0 = 100\;\mathrm{erg\ cm\ s}^{-1}$ as a characteristic scale at which line driving becomes significantly suppressed and $\alpha_{\xi} = 2$ to represent a smooth but relatively rapid transition in the suppression of the wind acceleration with increasing ionization.
Although the exact value of $\alpha_{\xi}$ is uncertain, the qualitative behavior reported here is expected to remain valid as long as the suppression occurs over a finite range of ionization parameter.
We adopt these values as our fiducial parameters. 
To examine the sensitivity of our results to the ionization threshold, 
we also perform additional simulations with $\xi_0 = 30\;\mathrm{erg\ cm\ s}^{-1}$ for selected models.

\section{Results}
\label{sec:result}

As mentioned above, our study here centers on four different models.
The first two are for studying the basic dynamical effect of the wind on the accretion flow: One with only the RLO and the other with the RLO and the wind. These models have no ionizing source. The other two are to investigate the effect of photoionization on the wind dynamics and resulting accretion flows: One with a strong ionization source with $L_\mathrm{X} = 10^{38}\;\mathrm{erg\ s}^{-1}$, and the other with a moderate ionization source with $L_\mathrm{X} = 10^{37}\;\mathrm{erg\ s}^{-1}$.
All simulations have run over $5 P_\mathrm{orb}$. Note that this run time is significantly shorter than the viscous timescale of the resulting accretion disks, but is long enough to examine the dynamical effect of the wind.
Table~\ref{tab:models} summarizes the main simulation parameters of these models.

\begin{table}
\caption{
Main simulation parameters.
}
\begin{tabular}{ccc}
\hline
Model $^{*}$ & \begin{tabular}{c}
        wind mass-loss \\
        rate ($M_\odot\;\mathrm{yr}^{-1}$)
        \end{tabular} 
      & \begin{tabular}{c}
        Ionization \\
        luminosity \\
        $L_\mathrm{X}$ ($\mathrm{erg}\;\mathrm{ s}^{-1}$)
        \end{tabular} \\
\hline
RLO only & 0 & 0 \\
RLO$+$wind & $4.18 \cdot 10^{-7}$ $^{**}$ & 0 \\
RLO$+$wind$+L_\mathrm{X}=10^{38}$ & $4.18 \cdot 10^{-7}$ $^{**}$ & $10^{38}$ \\
RLO$+$wind$+L_\mathrm{X}=10^{37}$ & $4.18 \cdot 10^{-7}$ $^{**}$ & $10^{37}$ \\
\hline
\end{tabular}\\
* In all models, the RLO mass-injection rate is fixed to $8.03 \times 10^{-9} M_\odot\;\mathrm{yr}^{-1}$. This rate corresponds to the X-ray luminosity $L_\mathrm{X} = 10^{38}\;\mathrm{erg\ s}^{-1}$ if all RLO injected gas steadily accretes onto the neutron star.\\
** Calculated by equation~(7) of \citet{Bjorklund2023} with
$M_{*} = 20 M_\odot$, $L_{*} = 3.05 \cdot 10^{5} L_\odot$, $T_\mathrm{eff} = 39,000\;\mathrm{K}$, 
and the solar metallicity. Here, $L_{*}$ is calculated by $L_{*} = 4 \pi R_{*}^2 \sigma T_\mathrm{eff}^4$ with $R_{*}=12\;R_\odot$.
\label{tab:models}
\end{table}

\subsection{Global flow structure}
\label{sec:flow}

We first describe the global morphology of the flow to provide a qualitative overview of the wind-stream interaction.

In what follows, we define the disk, or more precisely the bound flow around the neutron star, 
as gas that is gravitationally bound to the neutron star and located inside 
its Roche lobe, irrespective of particle origin.

\begin{figure}
\begin{center}
\includegraphics[width=0.23\textwidth]{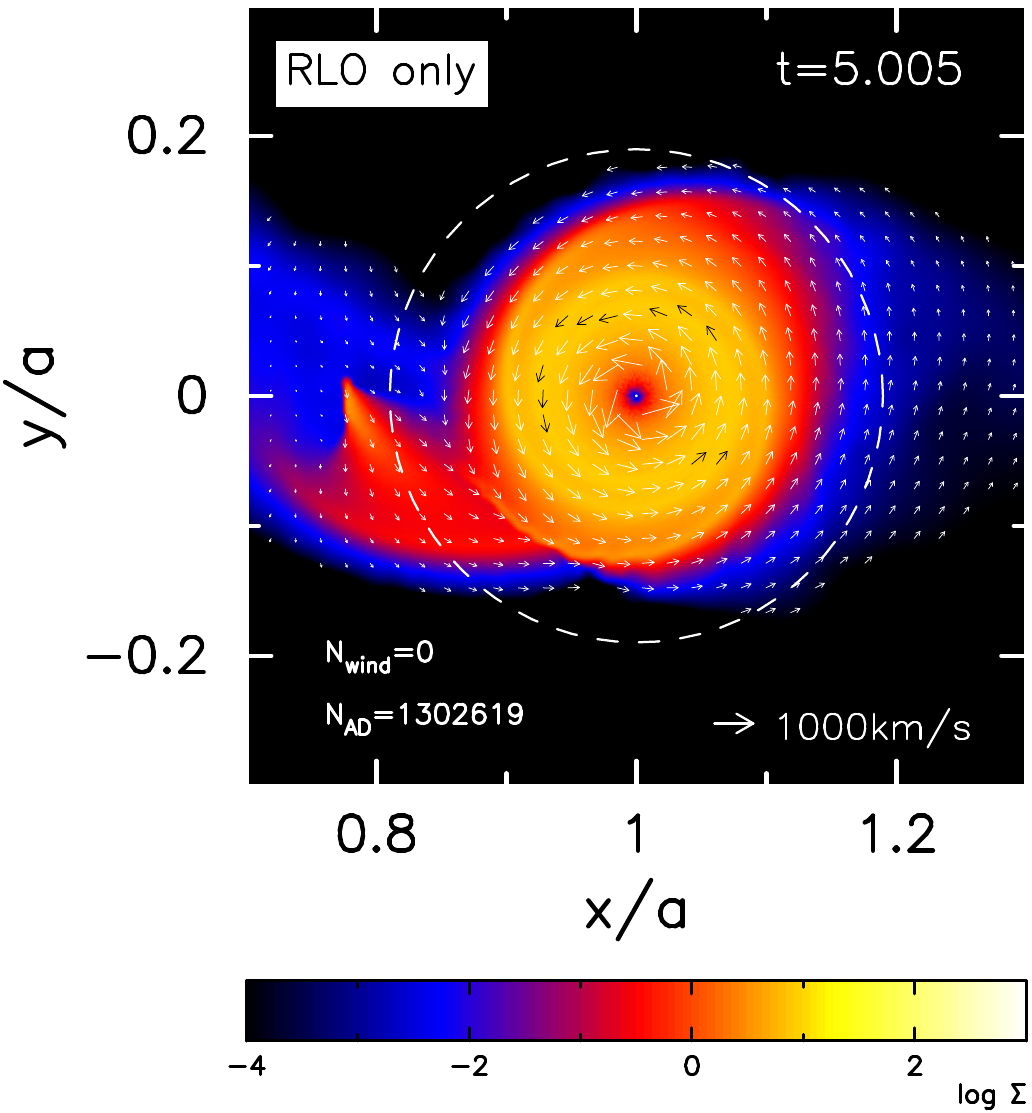}
\includegraphics[width=0.23\textwidth]{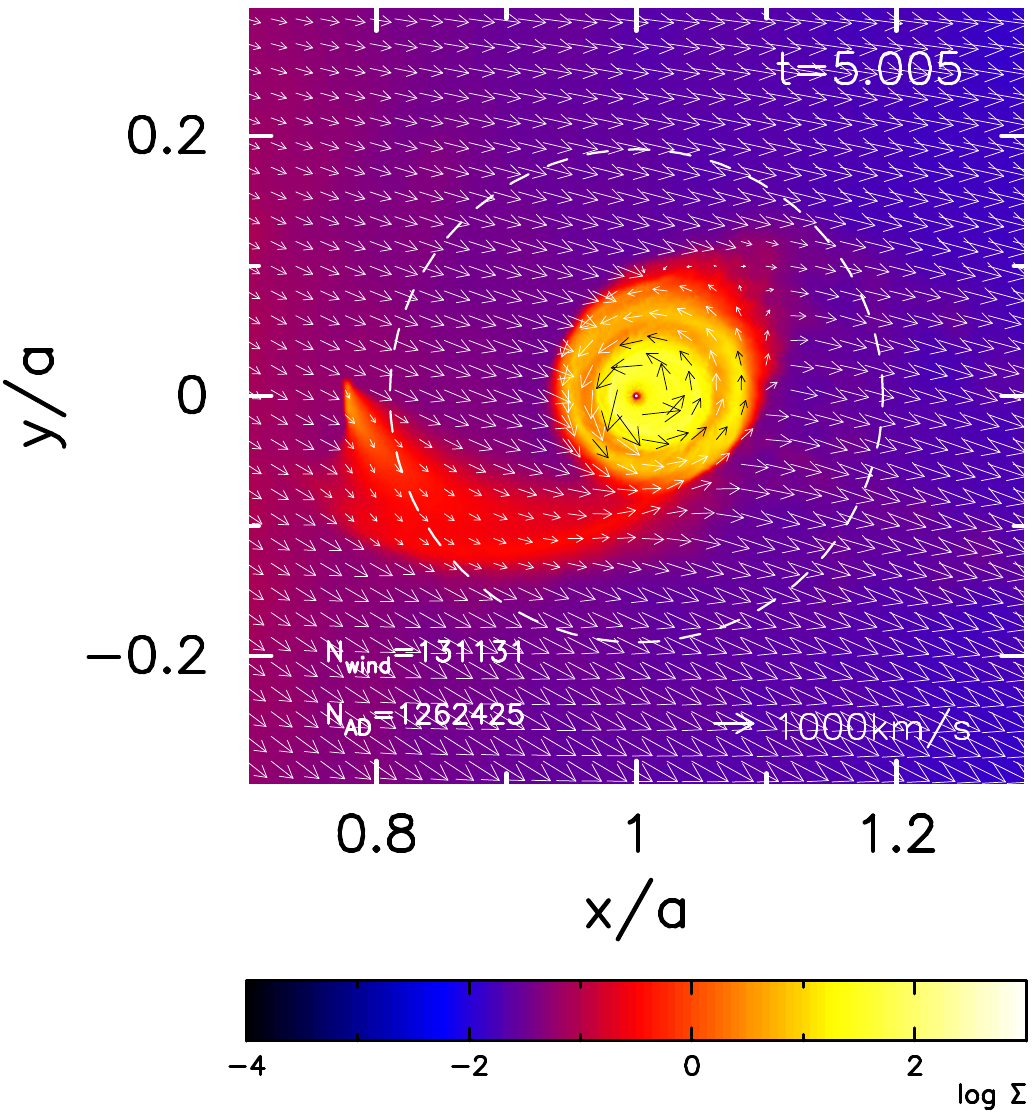}\\
\includegraphics[width=0.23\textwidth]{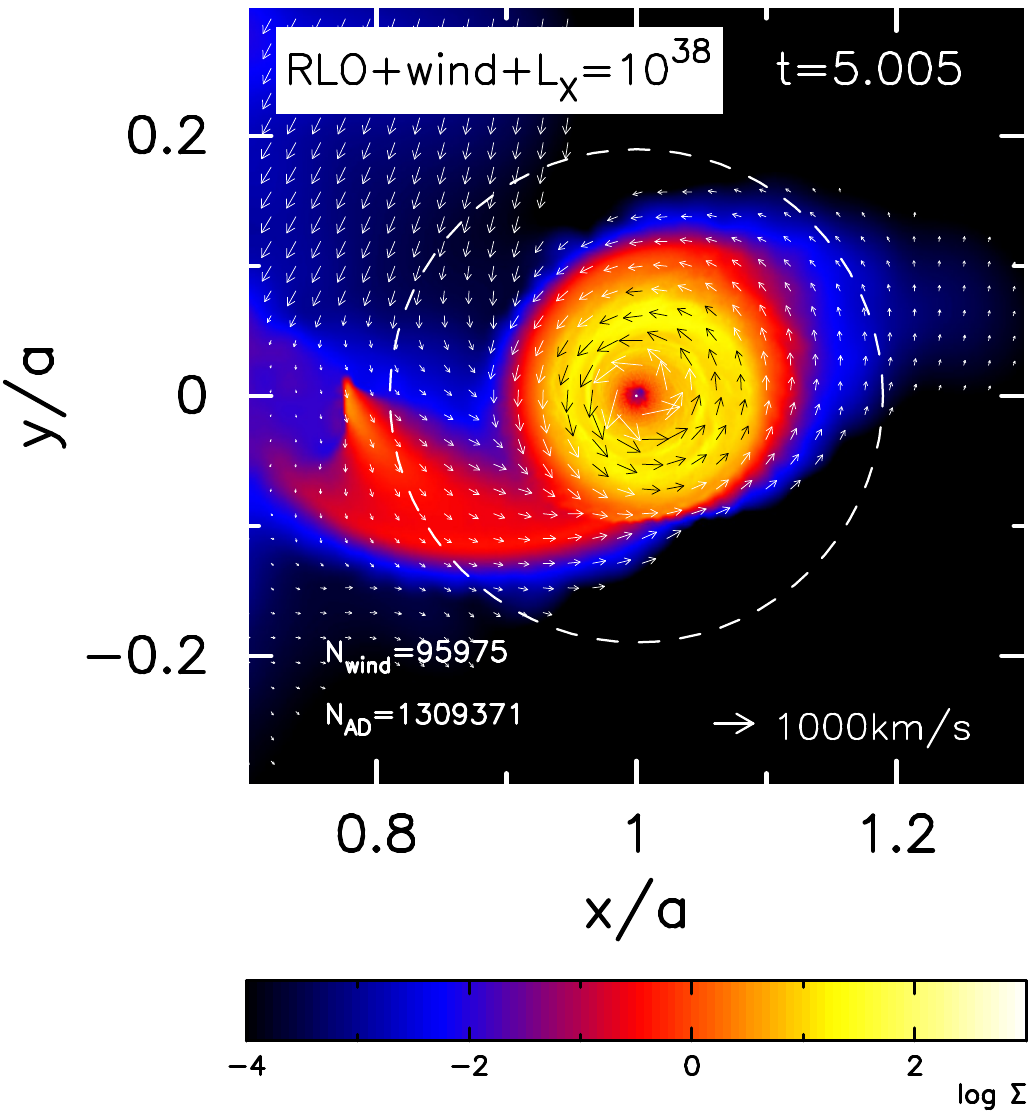}
\includegraphics[width=0.23\textwidth]{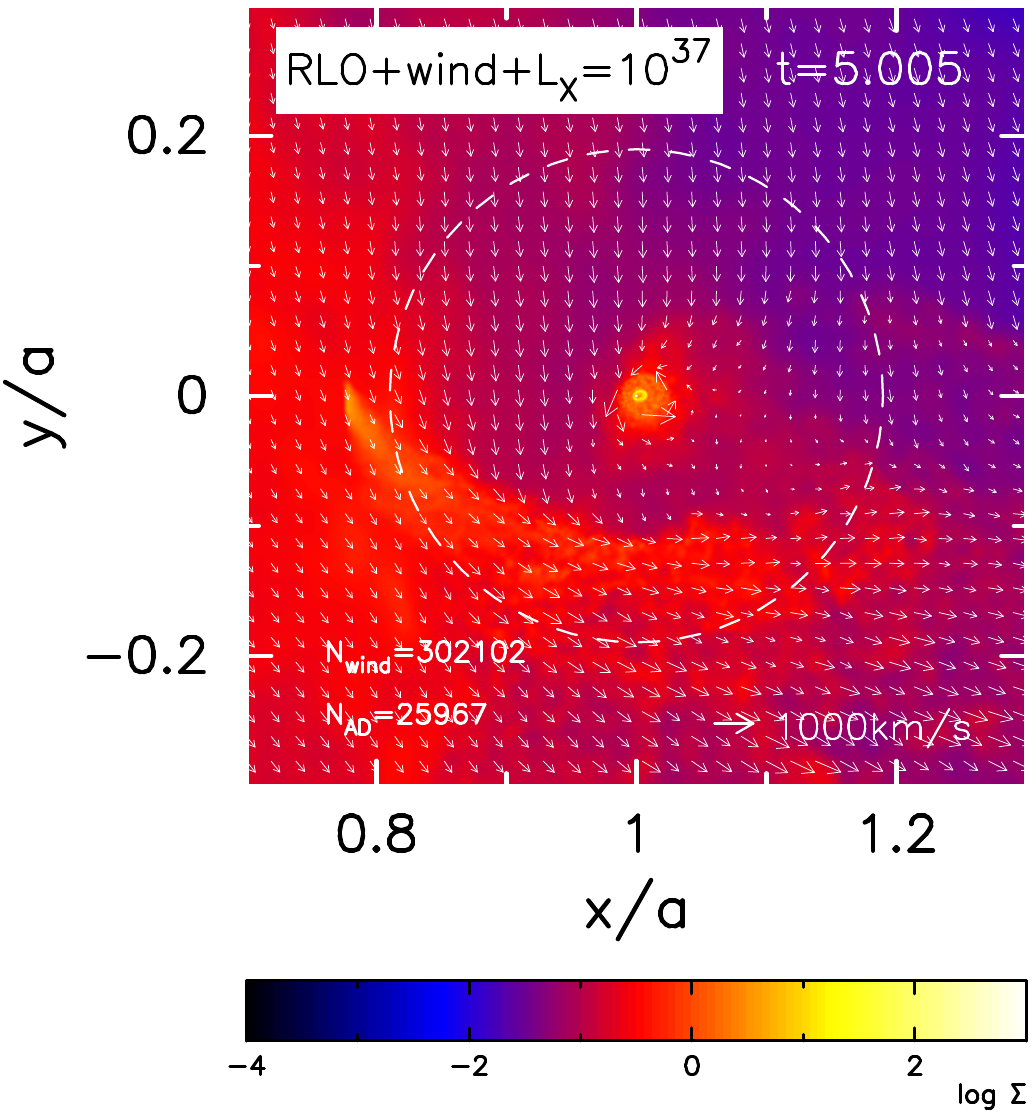}
\end{center}
\caption{
Surface density distributions and velocity fields at the end of the run. The panels are for (top left) RLO without stellar wind, (top right) RLO with stellar wind, (bottom left) RLO with stellar wind and strong ionization luminosity of $10^{38}\,\mathrm{erg\ s}^{-1}$, and (bottom right) RLO with stellar wind and moderate ionization luminosity of $10^{37}\,\mathrm{erg\ s}^{-1}$. For readability, the color of velocity arrows is changed between bright and dark backgrounds. In each panel, time is normalized by $P_\mathrm{orb}$ and $N_\mathrm{wind}$ and $N_\mathrm{AD}$ are the number of wind particles and that of RLO particles bound to the neutron star and inside its Roche lobe (white dashed line).
{Alt text: Four surface-density maps with velocity vectors at the end of the simulations. The RLO-only model forms an extended disk around the neutron star, the non-ionized wind model forms a smaller disk, the strongly ionized model resembles the no-wind case, and the moderately ionized model shows a compact central flow with the RLO stream strongly displaced by the wind.}}
\label{fig:sigma+v}
\end{figure}

Figure~\ref{fig:sigma+v} shows a snapshot at the end of each simulation run ($t = 5 P_\mathrm{orb}$). In each panel, the surface density distribution is plotted in the logarithmic scale, and the arrows overlayed on the surface density display the velocity field. 
In the 'RLO only' case, an accretion disk first forms near the circularization radius, 
i.e., the radius at which the specific angular momentum of the transferred material equals that of a circular Keplerian orbit around the compact object (for further detail, see Section 4.5 of \cite{Frank2002}),
and gradually expands outward as angular momentum is redistributed. Although the disk is still growing at the end of the run, it will finally be settled at a state where the viscous torque balances with the tidal/resonant torque.

On the other hand, when the RLO occurs in the strong wind environment ('RLO + wind' case) with no photoionization effect, the wind is efficiently accelerated and fills a large volume as a global
background flow. Although an accretion disk similarly forms as in the no-wind case, the disk size is significantly smaller, suggesting that the dynamical effect of the wind is the angular momentum removal from the disk. We will discuss this effect in detail in a later section.

When the ionization effect is taken into account, the flow structure sensitively depends on the strength of the ionizing source. In the case of $L_\mathrm{X} = 10^{38}\;\mathrm{erg\ s}^{-1}$ ('RLO + wind + $L_\mathrm{X}=10^{38}$' case), the wind acceleration is strongly suppressed. 
Most injected wind particles fail to escape and fall back onto the donor star (``failed wind'').
As a result, the effective wind density near the compact object is reduced,
and the RLO stream circularizes efficiently, forming an accretion disk of the size comparable with that without wind and significantly larger than in the case of no ionization.

For a moderate $L_\mathrm{X} = 10^{37}\;\mathrm{erg\ s}^{-1}$ ionization, the flow structure drastically changes. The wind acceleration is suppressed, so that it becomes a significantly denser wind, strongly pushing the RLO stream away from the neutron star. Consequently, most of the RLO particles escape the system and only a small fraction can enter orbits bound to the neutron star, forming a tiny accretion disk.

\begin{figure}
\begin{center}
\includegraphics[width=0.365\textwidth]{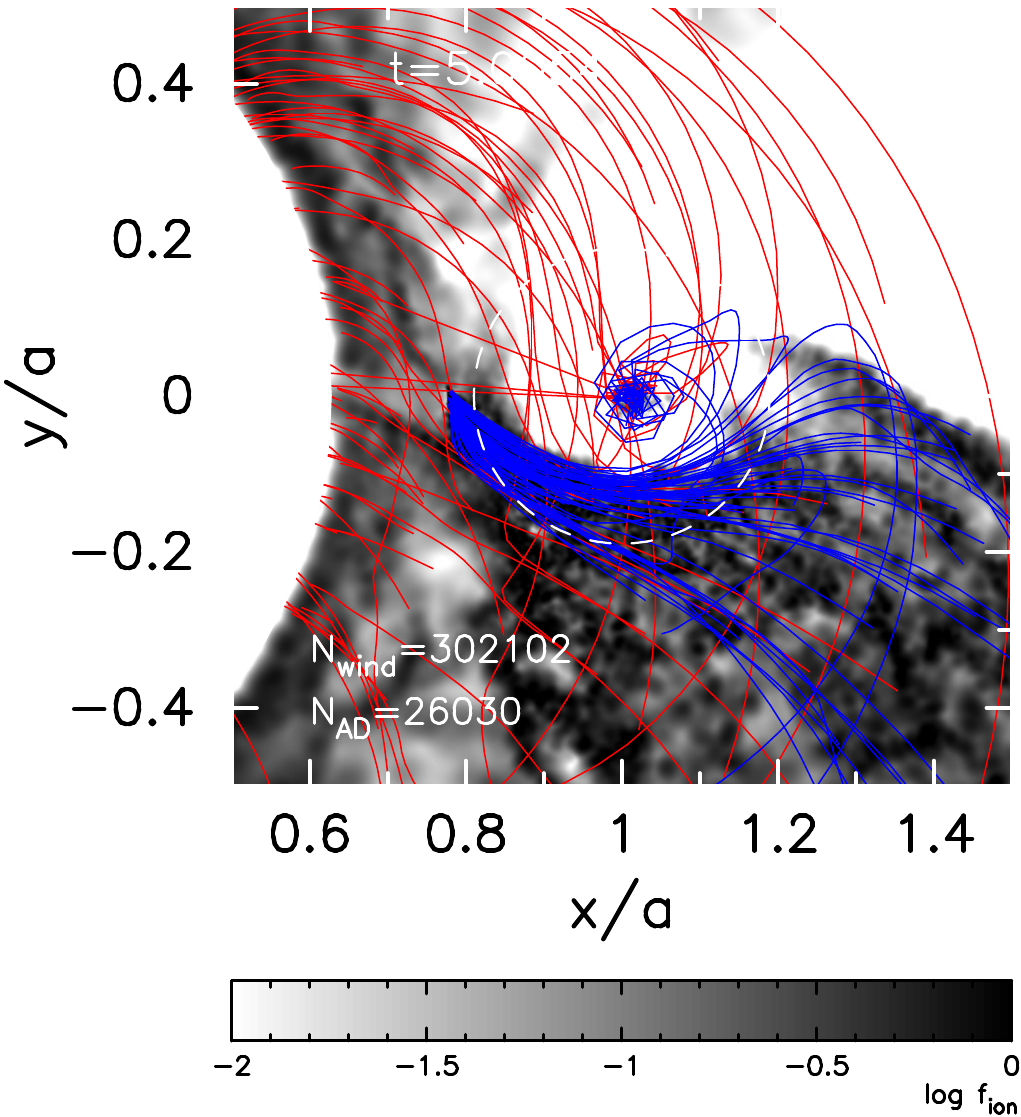}
\end{center}
\caption{
Sample trajectories of selected RLO-origin and wind-origin particles in the moderate-ionization model (RLO+wind+$L_\mathrm{X}=10^{37}\;\mathrm{erg\;s}^{-1}$) at $t = 5\;P_\mathrm{orb}$. The blue lines denote the trajectories of RLO particles projected onto the orbital plane, while the red lines are those of wind particles. Shown in the background is the reduction factor, $f_\mathrm{ion}$, of wind acceleration in the orbital plane.
{Alt text: Sample particle trajectories in the moderate-ionization model. Blue trajectories show RLO-origin particles and red trajectories show wind-origin particles projected onto the orbital plane, over a background map of the wind-acceleration reduction factor. The wind is bent toward the compact object in the strongly ionized region and collides with the RLO stream, pushing much of the RLO material away from the neutron star.}}
\label{fig:streamlines}
\end{figure}

This wind-RLO interaction is vividly illustrated in Figure~\ref{fig:streamlines}, which shows sample trajectories of RLO-origin and wind-origin particles at $t = 5\;P_\mathrm{orb}$. The blue lines denote the stream lines of RLO particles projected onto the orbital plane, while the red lines are those of wind particles. Shown in the background is the reduction factor, $f_\mathrm{ion}$, of wind acceleration in the orbital plane. We can see from Figure~\ref{fig:streamlines} that the upstream wind flow is bent in the $f_\mathrm{ion} \ll 1$ region toward the compact object, that the bent wind flow  collides with the RLO stream over the whole length, and that most of the RLO particles are pushed away from the system.

\begin{figure}
\begin{center}
\includegraphics[width=0.365\textwidth]{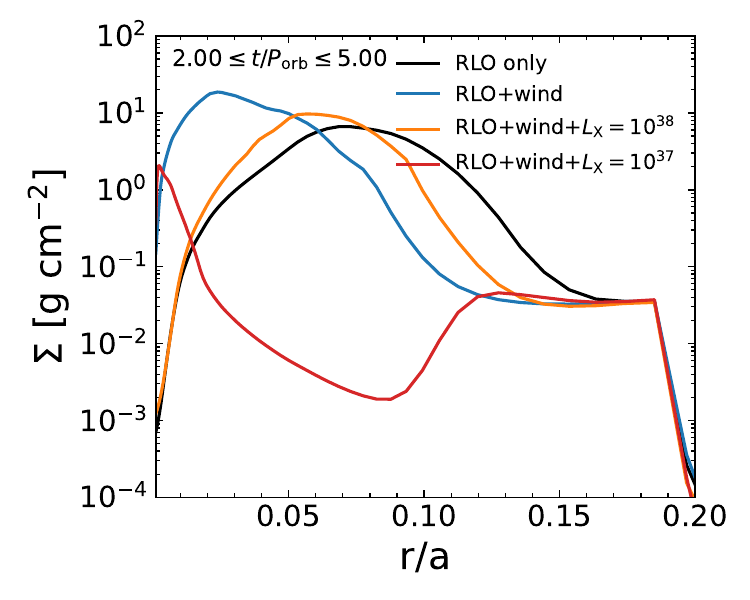}\hspace*{2mm} \\
\includegraphics[width=0.35\textwidth]{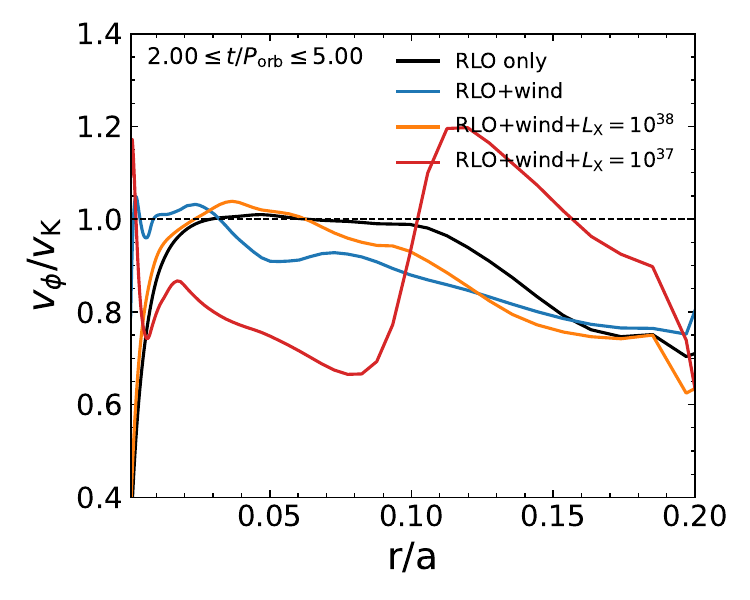} \\
\includegraphics[width=0.375\textwidth]{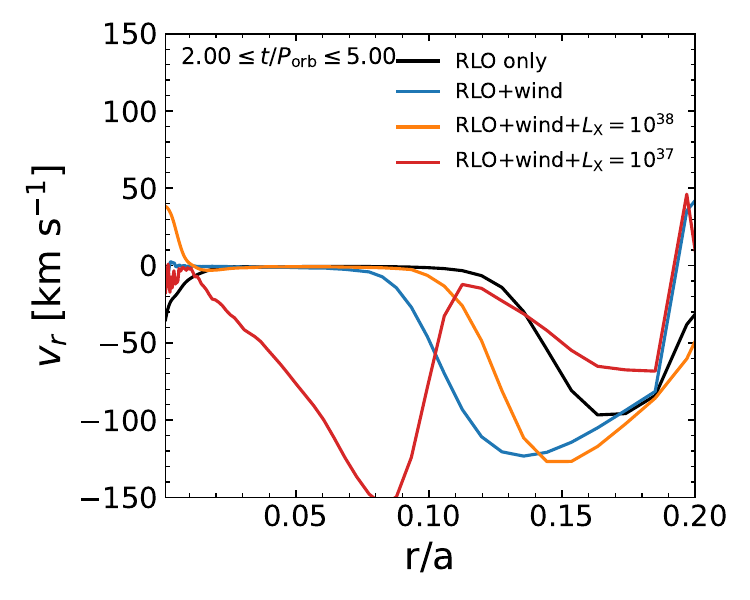}\hspace*{4mm}
\end{center}
\caption{
Radial profiles of azimuthally averaged disk quantities for the four cases, averaged over $2\le t/P_\mathrm{orb} \le 5$. From top to bottom, the surface density, the azimuthal velocity normalized by the local Keplerian velocity, and the radial velocity. In the bottom panel, negative velocities mean inflows.
{Alt text: Time-averaged radial profiles of disk quantities for the four models. The panels show surface density, azimuthal velocity relative to the Keplerian value, and radial velocity. The no-wind and strongly ionized models show extended, nearly Keplerian disk structures, the non-ionized wind model has a smaller disk, and the moderately ionized model shows a compact, rapidly inflowing, sub-Keplerian structure.}}
\label{fig:rad_prof}
\end{figure}

Figure~\ref{fig:rad_prof} compares the radial disk structure of the four cases averaged over $2\le t/P_\mathrm{orb} \le 5$. From top to bottom, the panels show the surface density, the azimuthal velocity normalized by the local Keplerian rotation velocity, and the radial velocity. The surface density is azimuthally averaged, while the velocity components are azimuthally and vertically averaged. In the bottom panel, $v_r < 0$ means inflow. 

From these panels, we observe that the disks of the no-wind and the strong ionization cases have similar structures, except that the latter is smaller because of (weak) wind effect. Given the surface peak near the disk outer radius, the azimuthal velocity close to Keplerian, and the small radial velocity all indicate that the disks of these no-wind and weak wind cases are controlled by viscosity, and are still evolving inward.

On the other hand, the disk in the case of strong wind without ionization ('RLO + wind' case) has a rather different disk structure: the surface density keeps increasing toward very small radii, the slightly sub-Keplerian azimuthal velocity, and small but noisy radial velocity suggest that the wind strongly affects the disk structure, not only making the disk size smaller, but also carrying disk gas inward much more rapidly than in a viscous disk.

In contrast to these three cases, the disk structure of the moderate ionization case shown in Figures~\ref{fig:sigma+v} and \ref{fig:rad_prof} suggests that it is in a different accretion regime.
The RLO stream is not collimated toward the neutron star, as in other models, but diffuses out beyond the Roche lobe, from which only a thin gas stream extends to a centrally condensed region near the neutron star (Figure~\ref{fig:sigma+v}). Unlike viscous Keplerian disks, the azimuthal velocity is significantly sub-Keplerian and the inward motion is supersonic. In other words, gas is almost free-falling with angular momenta much smaller than in the no- or strong-ionization cases, a characteristic seen in wind accretion.

\subsection{Transition from RLO-dominated to wind-dominated accretion}
\label{sec:transition}

\begin{figure}
\begin{center}
\includegraphics[width=0.46\textwidth]{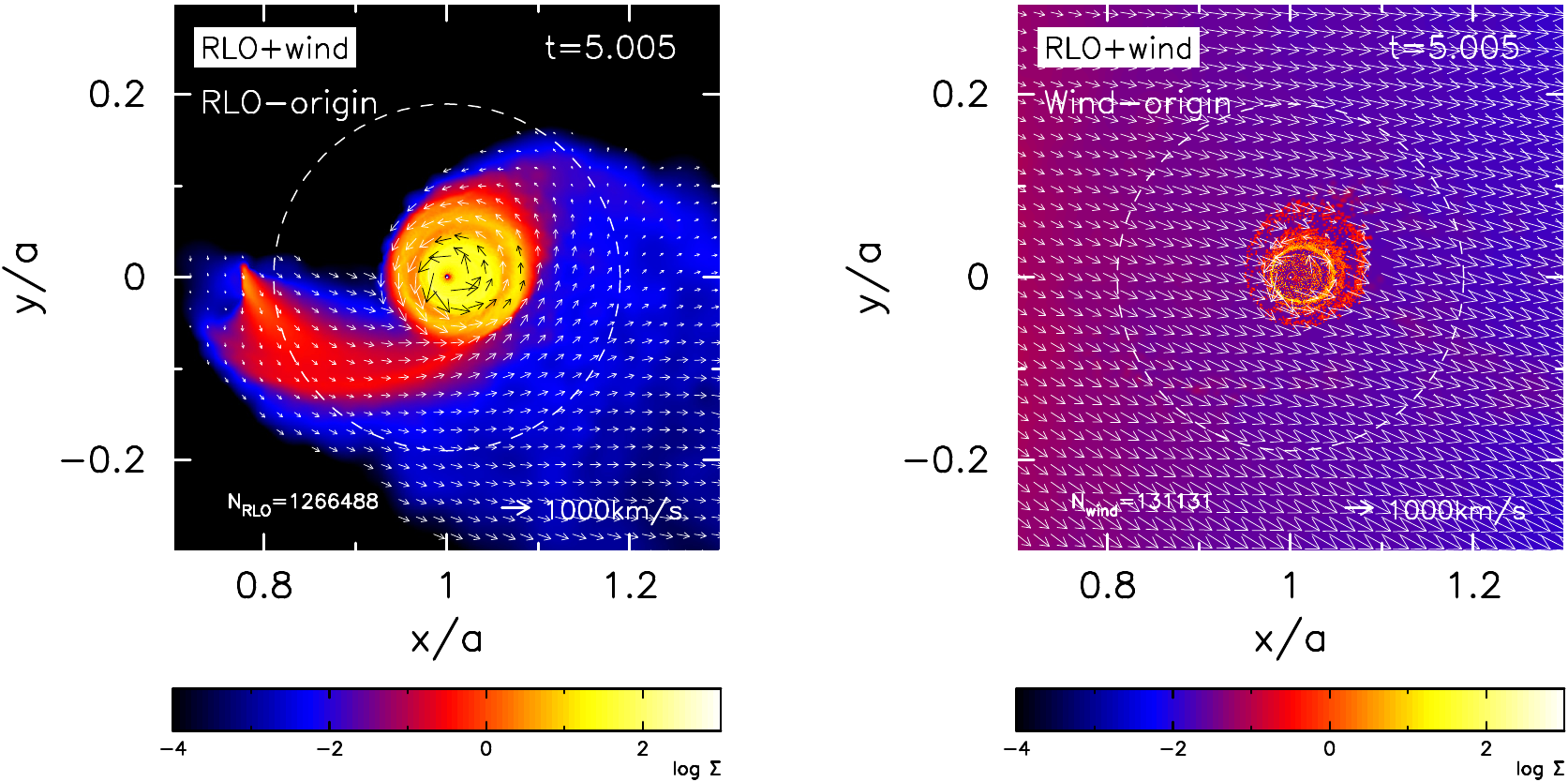} \\
\includegraphics[width=0.46\textwidth]{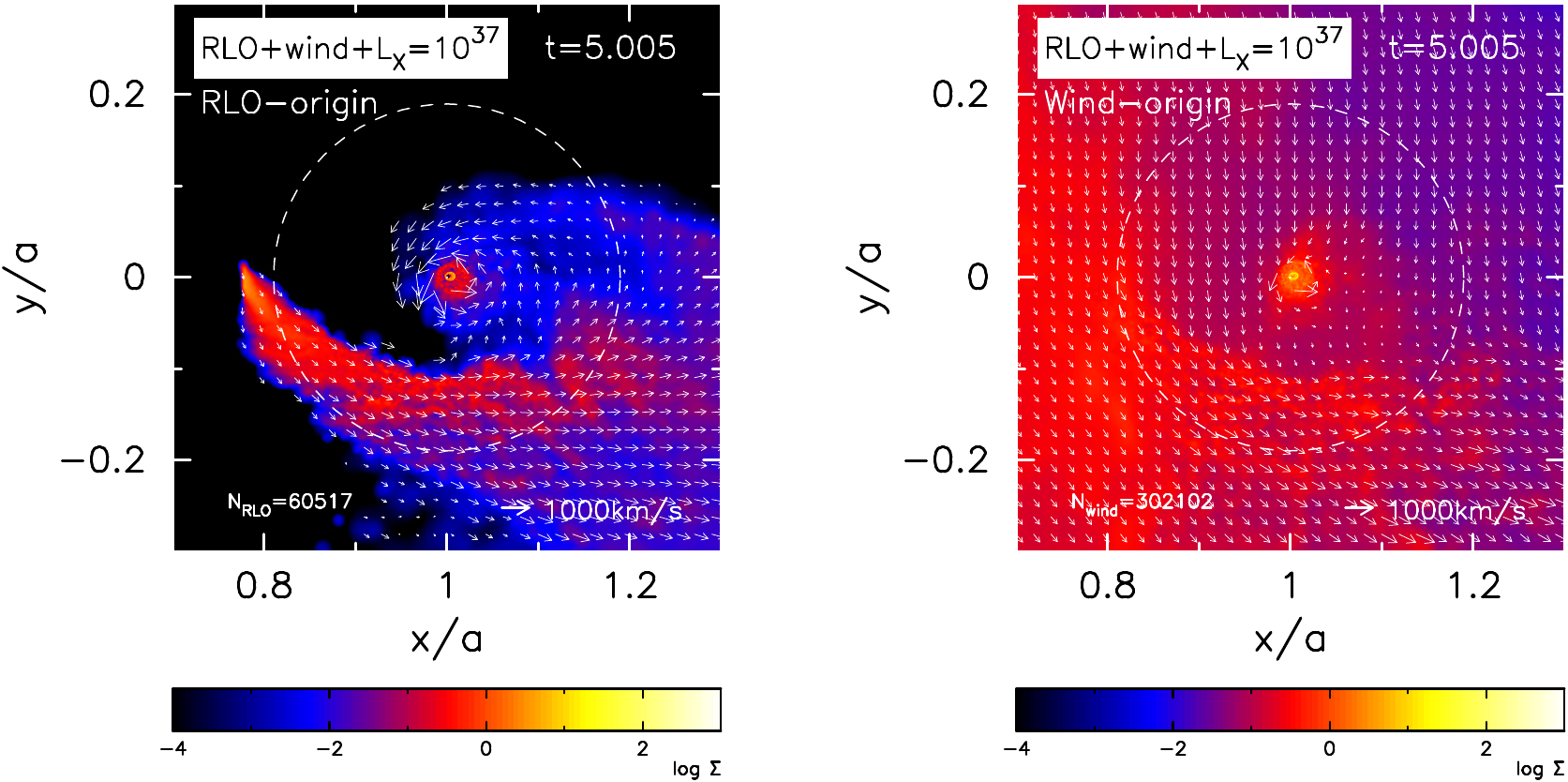}
\end{center}
\caption{
Surface density distribution and velocity field at the end of the run: (left) the RLO-origin particles and (right) wind-origin particles. The top panels are for the non-ionized wind case, while the bottom panels are for the moderate-ionization case. The other format of the figure is the same as that of Figure~\ref{fig:sigma+v}.
{Alt text: Origin-separated surface-density maps and velocity fields for the non-ionized and moderately ionized wind models. In the non-ionized model, RLO-origin gas dominates the disk while wind-origin gas contributes near the neutron star. In the moderately ionized model, much of the RLO-origin material remains in an extended stream, whereas wind-origin gas dominates the central region around the neutron star.}}
\label{fig:sepcomp}
\end{figure}

In order to clarify the nature of the accretion flow in the moderate-ionization case, we have examined the flow structure of the RLO-origin particles and the wind-origin particles, separately.

Figure~\ref{fig:sepcomp} presents the same snapshots as in Figure~\ref{fig:sigma+v} as separate distributions of (left) the RLO-origin and (right) wind-origin particles. The top panels are for the non-ionization wind ('RLO + wind') case, while the lower panels are for the moderate-ionization ('RLO + wind + $L_\mathrm{X}=10^{37}$') case. The figure shows a few remarkable features. First, even in the non-ionized wind case, wind particles are captured in an inner region of the accretion disk, significantly contributing to the disk density near the neutron star. Second, in the moderate-ionization case, the wind is so strong that it pushes most of the RLO gas away from the system. As a result, the wind density dominates the density of RLO particles inside the Roche lobe. The wind-origin gas forms a larger-scale central condensation than the RLO-origin gas.

\begin{figure}
\begin{center}
\includegraphics[width=0.45\textwidth]{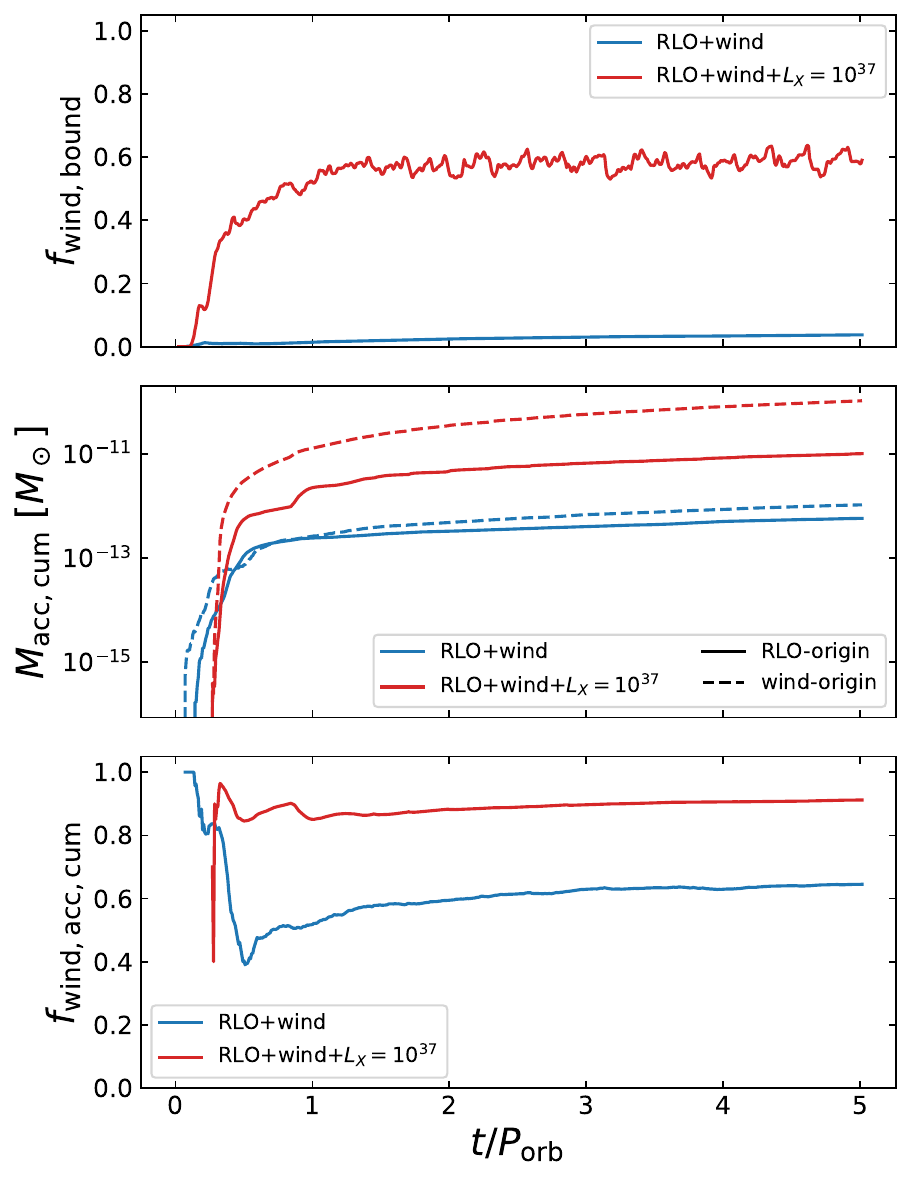}
\end{center}
\caption{
Evolution of the composition of bound and accreted material. 
Blue and red curves correspond to the RLO+wind and 
RLO+wind+$L_\mathrm{X}=10^{37}\ \mathrm{erg\ s^{-1}}$ models, respectively. 
Top: wind fraction of gas bound within the neutron star's Roche lobe. 
Middle: cumulative accreted mass; solid and dashed curves denote 
RLO-origin and wind-origin material, respectively. 
Bottom: cumulative wind fraction of the accreted material. 
Moderate X-ray ionization enhances the contribution of wind material both 
in the bound reservoir and in the accreted flow.
{Alt text: Time evolution of the wind-origin contribution to bound and accreted material. The non-ionized model has only a small wind-origin fraction in the bound reservoir but a substantial wind-origin fraction in the cumulative accreted mass. The moderately ionized model has a much larger wind-origin fraction both in the bound material and in the accreted material, showing a transition toward wind-dominated accretion.}}
\label{fig:wind_vs_RLO}
\end{figure}

To quantify this change in the accretion mode, we examine the composition of gas bound to the neutron star and that of accreted particles. Figure~\ref{fig:wind_vs_RLO} shows the evolution of the wind-origin fraction in the bound reservoir and in the cumulative accreted mass. 
Blue and red curves correspond to the no-ionization and 
moderate ionization models, respectively. From the figure, the following ionization-dependent characteristics are derived:
\begin{itemize}
\item In the no-ionization simulation, the bound flow around the neutron star is composed mostly of RLO-origin particles, with wind-origin particles contributing only a few percent (top panel). As shown in the middle panel, however, even in such an extreme composition, wind-origin particles are the majority ($\sim 60 \%$ from the bottom panel) in the cumulative accreted mass by the end of the simulation. Together with the snapshot in Figure~\ref{fig:sepcomp}, this implies that a part of wind-origin particles are captured in the inner region and accreted rapidly.

\item In the moderate ionization simulation, the RLO-origin and wind-origin particles have comparable masses in the bound reservoir (top panel). However, as shown in Figure~\ref{fig:sepcomp}, most of the RLO-origin material inside the Roche lobe remains in the extended stream, whereas the central region near the neutron star is dominated by wind-origin gas. Consequently, the cumulative accreted material is also dominated by wind-origin gas ($\sim 90 \%$ from the bottom panel). Thus, in the moderate ionization environment, the system is essentially fed by the donor wind.
\end{itemize}

Because our simulations follow the system for only several orbital periods, 
they are designed to capture the dynamical wind-stream interaction rather 
than the long-term viscous evolution of the accretion disk. 
On timescales longer than the viscous time, RLO-origin material stored in 
the disk may eventually dominate the accretion onto the neutron star in no- or weak-ionization environments.
Determining whether the accretion becomes RLO-dominated at later times 
requires longer simulations and is left for future work.

\subsection{Angular-momentum input by the donor wind}
\label{sec:wind_am}

We quantify how the donor wind supplies angular momentum to the accretion flow, which can act either as a source or a sink depending on the flow configuration.

We first compare the accretion flow formed with and without the donor
wind, without including X-ray photoionization. This comparison isolates
the dynamical effect of the stellar wind on the RLO-fed accretion flow.

We define the disk as the gas bound to the compact object within its Roche
lobe. For each snapshot, we compute the disk mass and angular momentum as
\begin{equation}
  M_\mathrm{disk} = \int 2\pi r \Sigma(r)\,dr,
\end{equation}
and
\begin{equation}
  J_\mathrm{disk} = \int 2\pi r \Sigma(r)\, r v_\phi(r)\,dr,
\end{equation}
where $r$ is the distance from the compact object projected into the orbital plane.
The mean specific angular momentum of the disk is then
\begin{equation}
  \bar{j}_\mathrm{disk} = \frac{J_\mathrm{disk}}{M_\mathrm{disk}}.
\end{equation}
The disk radius, $r_\mathrm{disk}$, is defined as the radius outside the
surface-density maximum where the surface density decreases most rapidly.

\begin{figure}
\begin{center}
\includegraphics[width=0.385\textwidth]{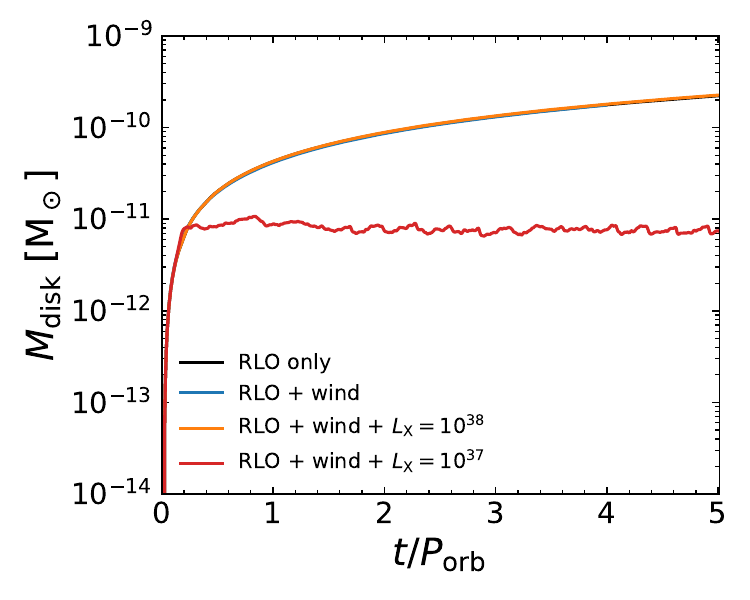}\hspace*{6mm} \\
\includegraphics[width=0.375\textwidth]{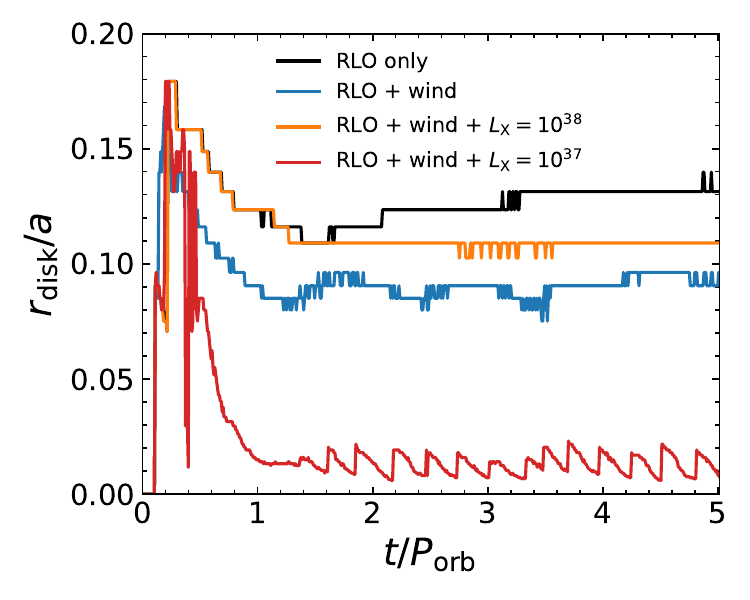}\hspace*{4mm} \\
\includegraphics[width=0.35\textwidth]{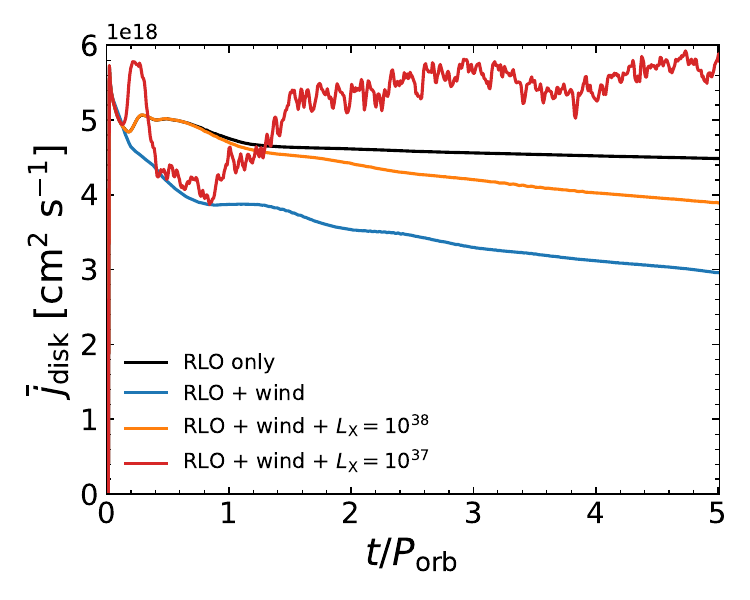}
\end{center}
\caption{
Time evolution of the disk mass (top), the disk radius (middle), and the mean specific angular momentum (bottom) of the disk for four cases shown in Figure~\ref{fig:sigma+v}. 
The $L_\mathrm{X} = 10^{37}$ case exhibits a
significant suppression of both disk mass and size compared to the other
three cases, indicating inefficient disk formation in the moderate
ionization regime.
{Alt text: Time evolution of disk mass, disk radius, and mean specific angular momentum for the four models. The no-wind, non-ionized wind, and strongly ionized models build up disk mass at similar rates, but the non-ionized wind model has a smaller disk radius and lower mean specific angular momentum. The moderately ionized model maintains a much smaller disk mass and radius but has a high mean specific angular momentum.}}
\label{fig:evol}
\end{figure}

Figure~\ref{fig:evol} shows that the disk masses in the wind and no-wind
cases and the strong ionization case grow almost linearly with time at nearly the same rate during the simulated interval. 
In contrast, the mean specific angular momentum is systematically smaller
in the non-ionized wind model. By the end of the calculation, the ratio
$\bar{j}_\mathrm{disk,wind}/\bar{j}_\mathrm{disk,no\,wind}$ decreases to
approximately 0.7 with $\bar{j}_\mathrm{disk, no wind} \sim 4.5 \times 10^{18}\;\mathrm{cm}^2\;\mathrm{s}^{-1}$ and $\bar{j}_\mathrm{disk, wind} \sim 3.0 \times 10^{18}\;\mathrm{cm}^2\;\mathrm{s}^{-1}$ (bottom panel). The disk radius is also smaller in the non-ionized wind model.
This indicates that the donor wind does not primarily suppress the
accumulation of RLO material, but instead modifies the angular-momentum
budget of the disk, removing the disk angular momentum rather than simply stripping disk gas.

Figure~\ref{fig:evol} also shows a distinctive behavior of the moderate ionization model. In addition to having a much smaller disk mass and radius than the other models, as we have already seen previously, the mean specific angular momentum, $\bar{j}_\mathrm{disk}$, is the largest among four models. It is remarkable that $\bar{j}_\mathrm{disk}$ of the moderate ionization case is even larger than of the no-wind case, where no wind torque is present. This suggests that in this case, the bound flow receives the angular momentum from the wind-origin particles, an opposite trend to the non-ionized wind model.

To identify the origin of the angular-momentum input, we measure the angular momentum carried by the inward-moving wind around the compact object. At a given radius $r_0$, we consider wind particles in a radial bin centered at $r_0$, and decompose their velocity in the frame centered on the compact object. For particles with inward radial velocity $(v_r<0)$, we define
\begin{equation}
  \dot{M}_\mathrm{w,in}(r_0,\phi)
  =
  \sum_{v_r<0} m_i v_{r,i},
\end{equation}
and
\begin{equation}
  \dot{J}_\mathrm{w,in}(r_0,\phi)
  =
  \sum_{v_r<0} m_i v_{r,i} j_{z,i},
\end{equation}
where $\phi$ is the azimuthal angle around the compact object and
\begin{equation}
  j_{z,i} = r_i v_{\phi,i}.
\end{equation}
Since $v_r<0$, $\dot{M}_\mathrm{w,in}$ is negative in this convention.
We therefore use the mass-flux-weighted specific angular momentum
\begin{equation}
  \bar{j}_\mathrm{w,in}
  = \frac{\dot{J}_\mathrm{w,in}}{\dot{M}_\mathrm{w,in}}
\end{equation}
to characterize the direction of the angular momentum carried by the
incoming wind. In our sign convention, the disk rotates in the positive
$j_z$ direction. Therefore, $\bar{j}_\mathrm{w,in}<0$ corresponds to a
retrograde wind component and hence to a negative torque on the disk.

\begin{figure*}
\begin{center}
\includegraphics[width=0.33\textwidth]{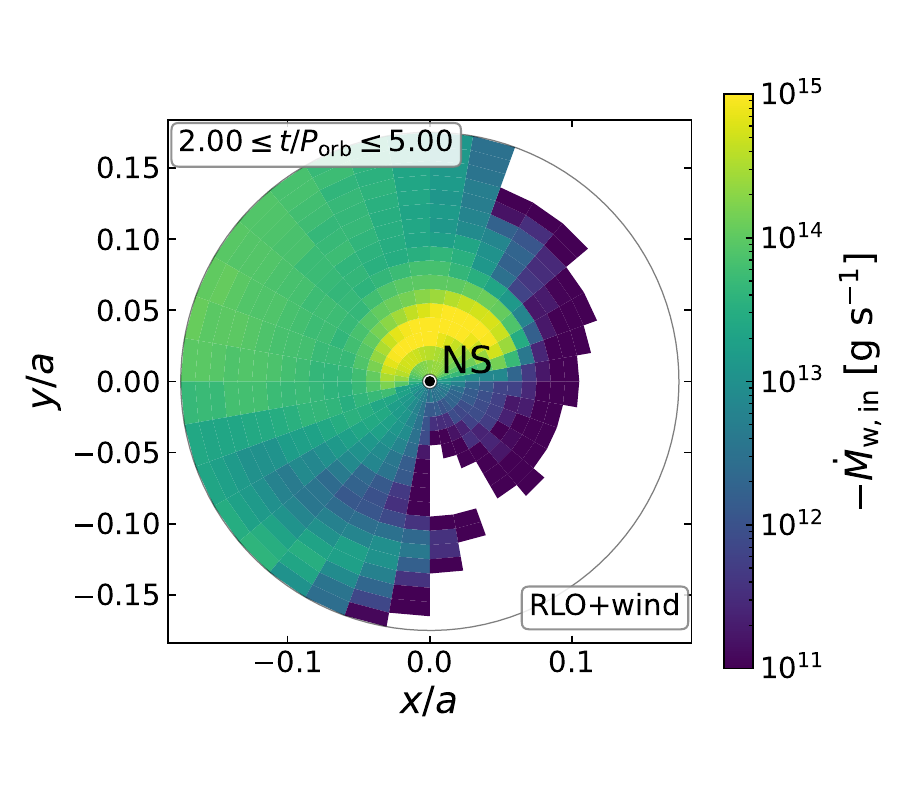}
\includegraphics[width=0.33\textwidth]{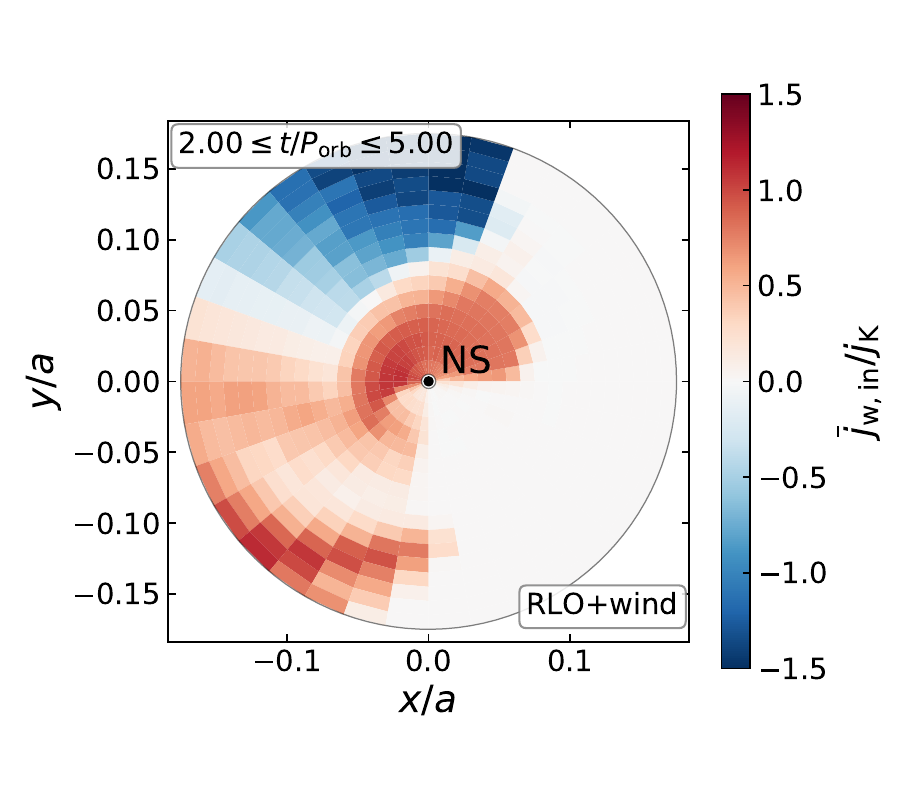}
\includegraphics[width=0.33\textwidth]{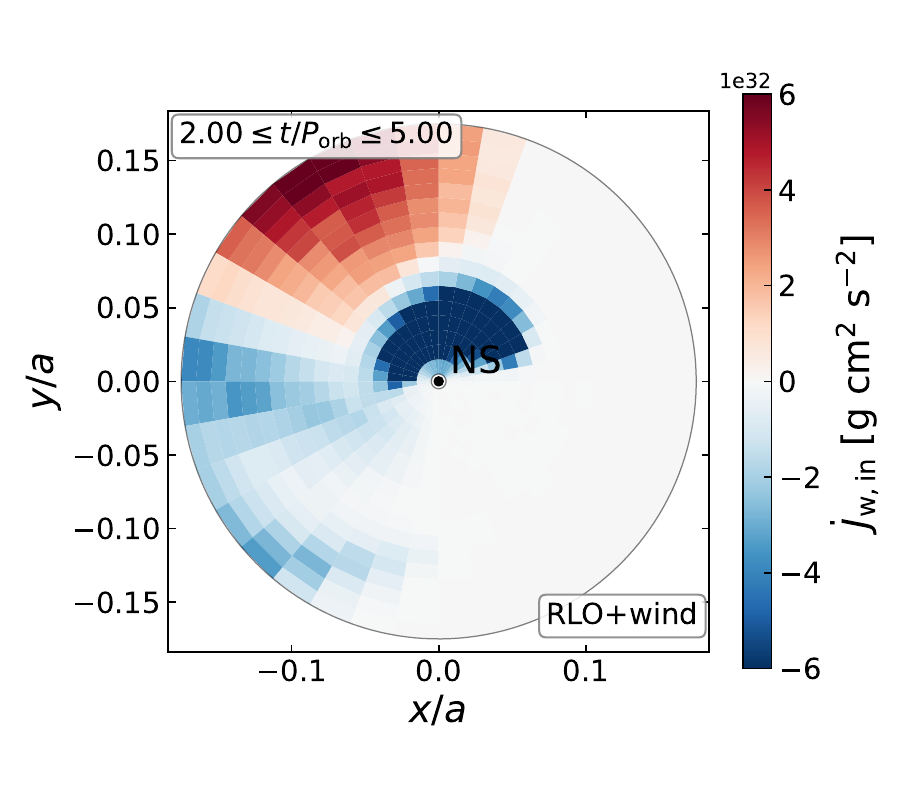}\\
\vspace*{-0.5cm}
\includegraphics[width=0.33\textwidth]{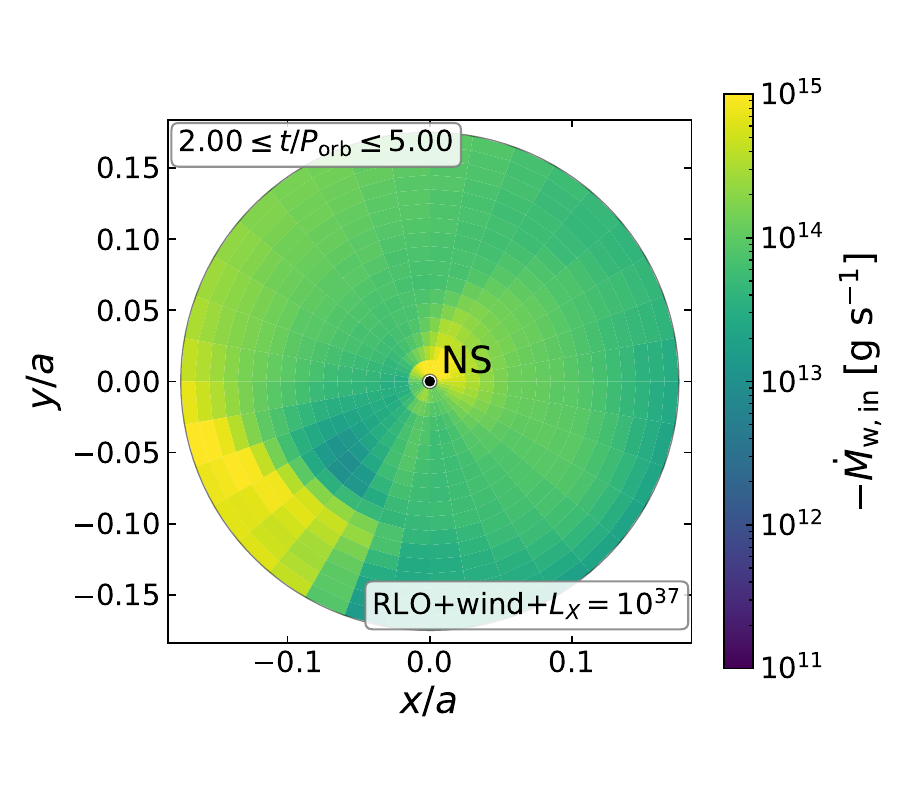}
\includegraphics[width=0.33\textwidth]{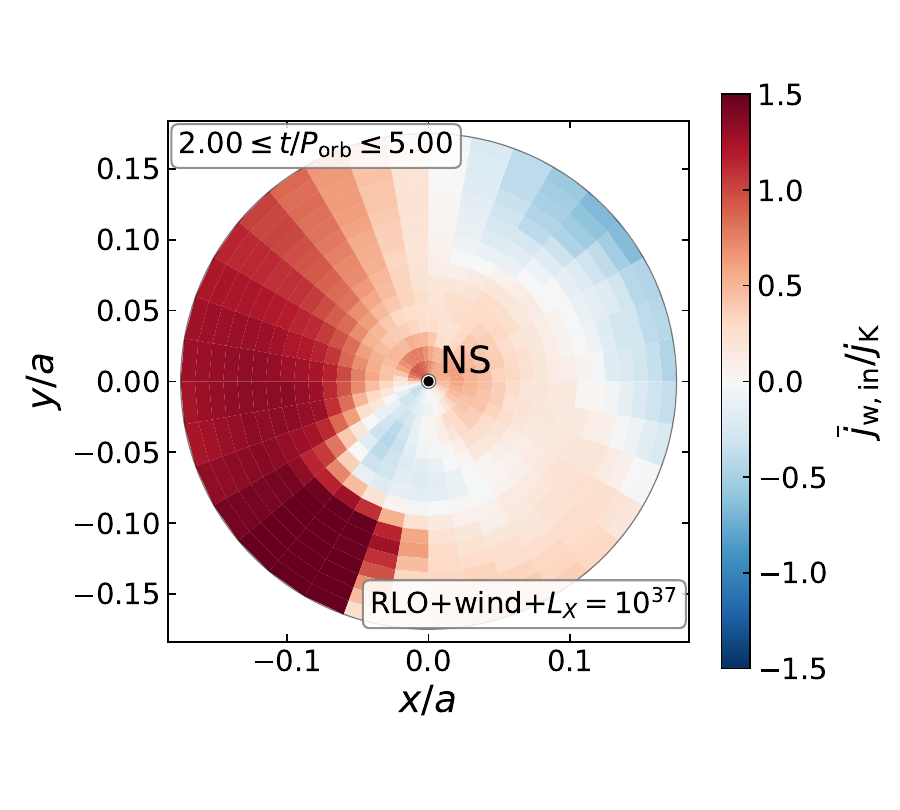}
\includegraphics[width=0.33\textwidth]{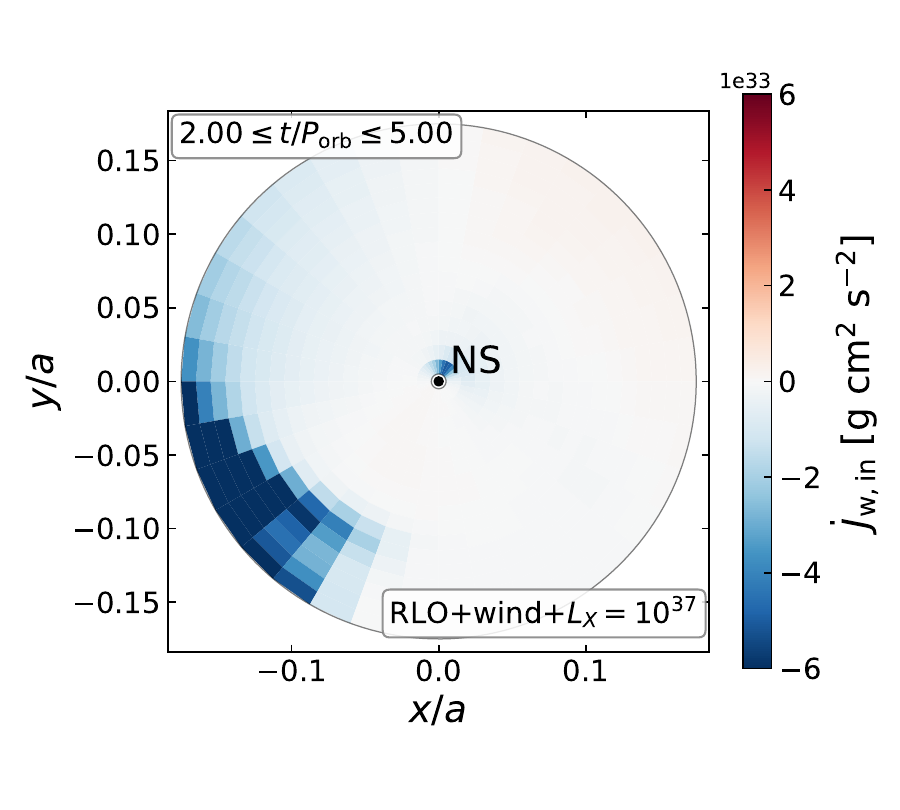}
\end{center}
\caption{
Time-averaged polar maps of the inward wind around the compact object
in the non-ionized wind model (upper panels) and the moderately-ionized 
($L_\mathrm{X}=10^{37}\ \mathrm{erg\ s^{-1}}$) wind model (lower panels). 
Each panel shows
(left) the inward wind mass flux $-\dot{M}_\mathrm{w,in}$,
(middle) the mass-flux-weighted specific angular momentum
$\bar{j}_\mathrm{w,in}/j_\mathrm{K}$, and
(right) the inward angular-momentum flux $\dot{J}_\mathrm{w,in}$.
The disk rotates counter-clockwise [in the positive $j_z$ direction], so negative
$\bar{j}_\mathrm{w,in}/j_\mathrm{K}$ indicates retrograde wind material.
Because $v_r<0$ for the selected particles, positive
$\dot{J}_\mathrm{w,in}$ corresponds to the inward transport of
retrograde angular momentum.
The color scale for $-\dot{M}_\mathrm{w,in}$ is logarithmic.
Only wind particles are included in this flux diagnostics; RLO particles
are excluded before the azimuthal binning. The white region in the inward mass-flux panel is where there is no inward particle.
{Alt text: Polar maps of inward wind flux diagnostics around the neutron star for the non-ionized and moderately ionized wind models. The non-ionized model shows a highly anisotropic inward wind, with retrograde angular momentum over much of the interaction region. The moderately ionized model shows broader inward wind flow and a strong prograde component aligned with the disturbed RLO stream.}}
\label{fig:polar}
\end{figure*}

Figure~\ref{fig:polar} shows the time-averaged polar distributions of
$\dot{M}_\mathrm{w,in}$, $\bar{j}_\mathrm{w,in}/j_\mathrm{K}$, and
$\dot{J}_\mathrm{w,in}$ around the compact object 
for the non-ionized wind model (upper panels) and the moderate ionization model (lower panels), 
where $j_\mathrm{K}$ is the Keplerian specific angular momentum given by
\begin{equation}
  j_\mathrm{K}(r) = \sqrt{G M_\mathrm{NS} r}.
\end{equation}

In the non-ionized wind model, the mass inflow is
strongly anisotropic. The inward wind carries retrograde specific angular
momentum over a substantial part of the interaction region (see the middle panel), but the
strongest angular-momentum flux does not occur simply where
$\bar{j}_\mathrm{w,in}/j_\mathrm{K}$ is most negative. Instead, it appears
where a large inward mass flux overlaps with retrograde specific angular
momentum. Thus, the wind torque is controlled by both the direction of
the angular momentum carried by the wind and the anisotropic mass flux
of the incoming wind. An elongated structure seen near the bottom of each panel is the place where the wind collides with the RLO stream. The wind torque depends on such a structure as well.

The mass inflow is also strongly anisotropic in the moderate ionization model.
However, the region of the strongest mass influx, which is along the RLO stream, now coincides with that of the maximum inflow of prograde angular momentum. This strongly indicates that in this model the wind exerts a strong positive torque on the RLO-origin particles.

\begin{figure}
\begin{center}
\includegraphics[width=0.45\textwidth]{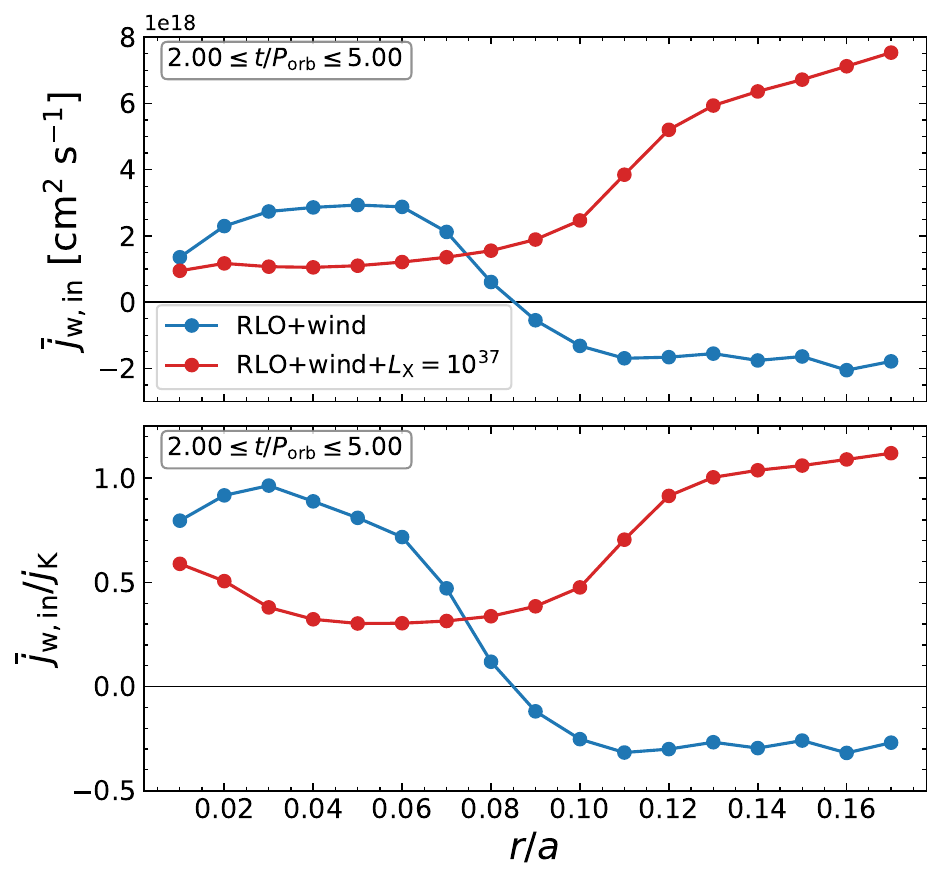}
\end{center}
\caption{
Radial profiles of the mean specific angular momentum carried by the inward wind around the compact object (upper panel) and its ratio to the local Keplerian value (lower panel) for the non-ionized wind model (blue lines) and the moderate ionization model (red lines).
{Alt text: Radial profiles of the specific angular momentum carried by the inward wind. In the non-ionized model, the inward wind is prograde at small radii but becomes retrograde near and outside the disk outer radius, indicating angular-momentum removal from the outer disk. In the moderately ionized model, the inward wind remains prograde over the full radial range shown.}}
\label{fig:windtorq}
\end{figure}

In order to characterize the radial dependence of the angular momentum carried by the inward wind, we azimuthally integrate the inward wind mass and angular-momentum fluxes shown in Figure~\ref{fig:polar}. In Figure~\ref{fig:windtorq}, the upper panel shows the mass-flux-weighted specific angular momentum carried by the inward wind around the compact object, whereas the lower panel shows its ratio to the local Keplerian value. The blue and red lines denote the non-ionized wind model and the moderate-ionization model, respectively.

In the non-ionized wind model, $\bar{j}_\mathrm{w,in}$ changes sign with radius. Around the outer disk radius, $r/a\sim0.1$, the inward wind carries retrograde angular momentum, with
$\bar{j}_\mathrm{w,in}/j_\mathrm{K} \sim -0.3$ or $\bar{j}_\mathrm{w,in} \sim -1.5\times10^{18}\ \mathrm{cm^2\,s^{-1}}$, comparable to the difference in $\bar{j}_\mathrm{disk}$ between the no-wind and non-ionized wind models. At smaller radii, however, $\bar{j}_\mathrm{w,in}$ becomes positive, which likely reflects captured wind-origin gas that has already interacted with the RLO stream and disk-like flow and is being advected inward. In contrast, in the moderate-ionization model, where the disk outer radius oscillates between $\sim 0.01 a$ and $\sim 0.02 a$, the inward wind carries prograde angular momentum over the full radial range shown. This prograde inflow acts mainly on the RLO stream, pushing it away from the neutron star and reducing the amount of RLO-origin material that can be accreted.

\subsection{Dynamical bifurcation of the RLO stream}
\label{sec:bifurcat}

\begin{figure}
\begin{center}
\includegraphics[width=0.45\textwidth]{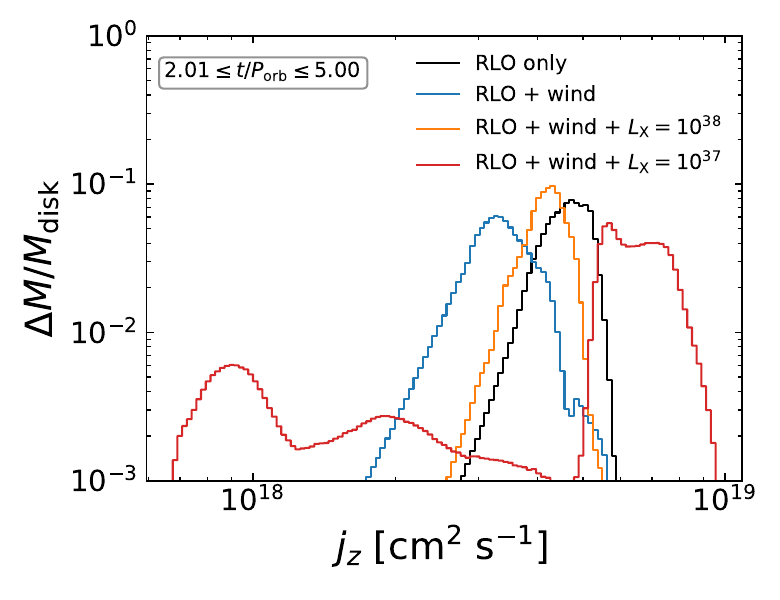}
\end{center}
\caption{
Time-averaged distributions of the specific angular momentum $j_z$
of bound RLO particles inside the neutron star's Roche lobe. The histograms are
averaged over $2 \leq t/P_\mathrm{orb} \leq 5$ and normalized by
the total disk mass in each model. The moderate ionization ($L_\mathrm{X}=10^{37}\ \mathrm{erg\ s^{-1}}$)
model shows a broad, bimodal distribution, indicating that the RLO
stream is dynamically split into low-$j_z$ disk-forming material and
high-$j_z$ material that follows a different trajectory.
{Alt text: Time-averaged distributions of specific angular momentum for bound RLO-origin particles inside the neutron star Roche lobe. The RLO-only, non-ionized wind, and strongly ionized models show relatively narrow, unimodal distributions. The moderately ionized model shows a broad bimodal distribution, indicating that the RLO stream splits into low- and high-angular-momentum branches.}}
\label{fig:jz_hist}
\end{figure}

To quantify the differences in flow behavior in more detail, we analyze the specific angular momentum distribution of RLO-origin particles within the Roche lobe of the compact object.
In Figure~\ref{fig:jz_hist}, we present time-averaged distributions of the specific angular momentum
$j_z$ of bound RLO particles inside the Roche lobe of the neutron star. The histograms are
averaged over $2 \leq t/P_\mathrm{orb} \leq 5$ and normalized by the total disk mass in each model.
In the non-ionized wind and strong ionization ($L_\mathrm{X}=10^{38}\;\mathrm{erg\ s}^{-1}$) cases, 
the distribution is approximately unimodal, indicating a single coherent flow component.
In contrast, the moderate ionization ($L_\mathrm{X}=10^{37}\;\mathrm{erg\ s}^{-1}$) case exhibits a clear bimodal distribution, consisting of high-$j_z$ and low-$j_z$ components.
This demonstrates that the anomalous behavior in the moderate ionization case
is not due to a uniform reduction of angular momentum, but rather due to
a splitting of the RLO flow into two distinct angular-momentum branches.

By mapping the spatial distribution of particles separated by a threshold
in $j_z$, we find that the low-$j_z$ branch corresponds to particles that penetrate
deep into the gravitational potential of the compact object, forming a
compact inner component and a narrow downstream flow.
On the other hand, the high-$j_z$ branch corresponds to particles that follow
larger-radius trajectories, forming an extended stream that partially
circulates but ultimately fails to fully circularize.
Importantly, the two branches are spatially separated, indicating that
the bimodality arises from a dynamical selection process rather than
a gradual redistribution of angular momentum.

We now investigate the physical origin of the bimodal angular momentum distribution identified above.
Time-dependent analysis shows that the splitting occurs near the compact
object when the RLO stream approaches the compact object.
At this stage, particles with smaller impact parameters lose angular momentum and
fall into the low-$j$ branch, while those with slightly larger impact parameters avoid strong deflection and remain in the high-$j$ branch.
Thus, the flow undergoes a dynamical bifurcation at the compact object.

The strength of this bifurcation depends sensitively on the wind structure:
Without ionization, the wind forms a relatively smooth global flow,
and the bifurcation is weak.
With strong ionization ($L_\mathrm{X}=10^{38}\;\mathrm{erg\ s}^{-1}$), however, 
the wind is largely suppressed via fallback,
and the RLO stream evolves nearly ballistically, again reducing the
bifurcation.
On the other hand, with moderate ionization ($L_\mathrm{X}=10^{37}\;\mathrm{erg\ s}^{-1}$), 
the wind is neither fully suppressed nor globally coherent. 
Instead, it moves as a significantly denser wind in the vicinity of the compact object.
This moderate regime maximizes the perturbation to the RLO stream,
leading to the strongest branch splitting.

\subsection{Sensitivity to the ionization threshold}
\label{sec:xi0_30_summary}

To examine the sensitivity of our results to the adopted ionization threshold, 
we performed an additional simulation with $\xi_0 = 30\;\mathrm{erg\ cm\ s}^{-1}$ for the $L_\mathrm{X} = 10^{37}\ \mathrm{erg\ s^{-1}}$ model. Below we summarize the results from this simulation. Detailed comparison with the case of $\xi_0 = 100\;\mathrm{erg\ cm\ s}^{-1}$ counterpart is given in Appendix~\ref{sec:xi0_30_result}.

In general, results for $\xi_0 = 30\;\mathrm{erg\ cm\ s}^{-1}$ are qualitatively the same as those for $\xi_0 = 100\;\mathrm{erg\ cm\ s}^{-1}$.
However, because of the stronger suppression of wind acceleration for $\xi_0 = 30\;\mathrm{erg\ cm\ s}^{-1}$, the wind becomes less disruptive than in the fiducial case of $\xi_0 = 100\;\mathrm{erg\ cm\ s}^{-1}$. Compared with the $\xi_0=100\;\mathrm{erg\ cm\ s}^{-1}$ case, the accretion disk is larger, while the disk mass is similar. The inward wind at radii outside the disk carries a larger prograde specific angular momentum, indicating that the incoming wind flow is more strongly biased toward the prograde direction. Nevertheless, the overall disturbance of the RLO stream and the bifurcation of the specific-angular-momentum distribution is weaker.

These results suggest that the disturbed-flow regime found at moderate 
ionization levels is robust against moderate changes in the ionization threshold, 
although higher-resolution calculations are required for quantitative convergence.

\section{Discussion}
\label{sec:discussion}

The results presented in Section~\ref{sec:result} suggest that the interaction between the stellar wind and the RLO stream fundamentally alters the accretion regime in RLO HMXBs.

\subsection{Physical picture of basic wind-RLO interaction}

The results presented in the previous section show that the donor wind is not merely a background component in RLO HMXBs, but can actively modify the accretion flow. In the non-ionized wind model, the RLO stream still forms a bound disk-like structure around the neutron star, but the donor wind interacts asymmetrically with the outer disk. Because the inward wind carries retrograde angular momentum over a substantial part of the interaction region, the disk loses angular momentum and becomes smaller than in the no-wind case.

This result suggests that the size of an RLO-fed disk in a RLO HMXB is not determined only by the circularization radius of the RLO stream or by viscous and tidal torques, but is affected by the negative torque exerted by the donor wind. The details of the interaction depend on the wind density and velocity around the compact object.

\subsection{Ionization-driven change of accretion regime}

A key result of this study is that X-ray ionization can qualitatively change the accretion regime. At high ionization luminosity of $10^{38}\;\mathrm{erg\ s}^{-1}$, the wind acceleration is strongly suppressed and most wind material falls back toward the donor. In this limit, little wind reaches the neutron star, and the accretion flow is similar to that of the no-wind case.

At moderate ionization luminosity ($L_\mathrm{X}=10^{37}\;\mathrm{erg\ s}^{-1}$), however, 
the donor wind is not fully suppressed but moves toward the compact object as a significantly denser and radially slower wind, which has the azimuthally dominated velocity component in the comoving frame with the binary.
Such a wind has a much larger impact on the RLO stream than in the no-ionization case and the strongly ionized failed-wind case, drastically changing the accretion flow structure and dynamics.
In this regime, the bound material around the neutron star becomes wind-rich, and the accreted material is dominated by wind-origin particles.

This demonstrates that the effect of ionization is fundamentally non-monotonic. The most disruptive regime can occur when the wind is partially suppressed but still reaches the  region around the compact object. In such a case, the wind can modify the RLO stream before it circularizes into a coherent disk and change the accretion regime from the RLO-fed to the wind-fed.

\subsection{Dynamical bifurcation of the RLO stream}

The moderate-ionization model also shows a broad, bimodal distribution of the specific angular momentum of bound RLO-origin particles. This suggests that the RLO stream undergoes a dynamical bifurcation near the neutron star. Part of the stream loses angular momentum and penetrates deep into the gravitational potential of the neutron star, while another part is pushed into a higher-angular-momentum branch.

This bifurcation is closely related to the wind-stream interaction. 
In the moderate-ionization model, the RLO stream first receives a positive torque from the dense prograde wind, increasing its specific angular momentum, and starts moving away from the compact object. Thus, the majority of the stream particles form the higher-angular-momentum branch. However, after the stream head wraps around the compact object counter-clockwise, it receives a negative torque from the retrograde wind. This is when the bifurcation occurs. The part well inside the Roche lobe loses its angular momentum and falls toward the compact object. The low angular momentum branch corresponds to this part. The remaining part, which already has a large enough angular momentum, also loses its angular momentum but manages to escape from the system.
Therefore, the suppression of disk formation is not simply caused by a uniform loss of angular momentum. Rather, it results from a restructuring of the RLO flow into multiple dynamical components.

\subsection{Astrophysical implications}

The results suggest that RLO HMXBs can exhibit accretion behavior that differs substantially from both ordinary RLO-fed systems and purely wind-fed systems. Depending on the ionization state of the donor wind, the system may transition between a disk-dominated RLO-fed regime and a wind-rich accretion regime.

Such transitions may be relevant to the observed variability in 
luminous HMXBs such as Cen X-3, SMC X-1, and LMC X-4, which show 
both relatively stable high-luminosity states and episodes of 
significant flux reduction \citep{Paul2005, RaichurPaul2008}. 
In these systems, X-ray photoionization is expected to modify 
the structure of the donor wind \citep{Stevens-Kallman1990}, 
potentially leading to changes in the accretion flow geometry.

In our model, moderate ionization leads to a wind-dominated 
accretion regime and suppresses disk formation despite a 
continuous RLO mass supply. This suggests that variability in 
the X-ray luminosity could induce transitions between different 
accretion regimes by altering the wind-stream interaction.

More generally, our results imply that RLO HMXBs should not be 
treated as simple high-mass analogues of LMXBs. The presence 
of a strong donor wind introduces an additional channel of 
angular-momentum exchange, which can fundamentally modify 
the structure and evolution of the accretion flow 
\citep{Martinez-Nunez2017}.

There are also systems such as Cyg X-1, where the donor nearly 
fills its Roche lobe while still driving a strong wind. In such 
systems, the wind-RLO interaction may play a key role in 
determining the accretion mode \citep{Gies2003}.

Recent XRISM observations of Cen X-3 have also revealed complex 
absorption structures and spectral variability associated with the 
circumstellar environment \citep{Mochizuki2024}. 
Given that the X-ray luminosity was $\sim 10^{37}\;\mathrm{erg\ s}^{-1}$ during the observation \citep{Mochizuki2024}, such features may be related to the disturbed wind-stream interaction 
suggested by our simulations, although detailed radiative-transfer 
calculations are required to establish a direct connection.

\subsection{Limitations of the present model}

Several limitations should be kept in mind. First, the wind-launching prescription is simplified. The wind acceleration is adjusted to emulate a beta-type velocity law, and the effect of ionization is included by reducing the wind acceleration force through a prescribed function of the ionization parameter. A more complete treatment would compute line driving and ionization-dependent wind launching self-consistently.

Second, the RLO mass injection method adopted in this study is also simplified. We injected gas particles from a small region located slightly outside the L$_1$ point, within a conical volume oriented toward the neutron star. Although improvements of the mass injection method will not qualitatively affect the major results obtained in this study, the implementation of a better method, such as in \citet{Dickson2024}, is desirable to have quantitatively robust results.

Third, the quantitative strength of the wind-stream interaction may depend on the adopted artificial-viscosity prescription. Although our prescription provides an effective Shakura-Sunyaev viscosity in disk-like regions, it acts primarily as a shock-capturing numerical dissipation in the wind-stream interaction region. Therefore, quantitative values such as the disk radius and the detailed strength of the wind-stream interaction should be interpreted with this uncertainty in mind.

Another limitation is that our simulations cover only several orbital periods. This is sufficient to study the dynamical interaction between the wind and the RLO stream, but not long enough to follow the long-term viscous evolution of the disk. In particular, in models where RLO-origin material is stored in the disk, the relative contribution of RLO-origin and wind-origin gas to the accreted material may change on longer timescales.
Long-term simulations are also needed to enable a more self-consistent treatment of the ionization effect. In this study, the X-ray luminosity is treated as an externally specified parameter. In a fully self-consistent model, however, $L_\mathrm{X}$ should be coupled to the instantaneous accretion rate onto the neutron star, introducing feedback between the accretion flow and the wind ionization state.
Because the viscous timescale of the disk is much longer than the duration of the present simulations, such feedback cannot be captured here. Over longer timescales, this coupling may alter the accretion flow structure and the relative contributions of RLO-origin and wind-origin material.

Finally, the moderate-ionization regime is likely sensitive to numerical resolution and to the adopted ionization prescription. Our additional calculations with a different ionization threshold suggest that the qualitative behavior is not fine-tuned to a single value of $\xi_0$ (see Appendix~\ref{sec:xi0_30_result}), but higher-resolution simulations are required for quantitative convergence.

\section{Conclusions}
\label{sec:conclusion}

We have investigated the dynamical interaction between the donor wind and the Roche-lobe overflow (RLO) stream in high-mass X-ray binaries using three-dimensional SPH simulations, including the effect of X-ray photoionization on the wind acceleration.

Our main results can be summarized as follows:

\begin{enumerate}

\item In the absence of X-ray ionization, the donor wind interacts asymmetrically with the accretion disk and carries retrograde angular momentum. This results in a net removal of angular momentum from the disk and leads to a significantly smaller disk compared to the no-wind case, while the disk mass growth rate remains similar.

\item X-ray ionization qualitatively changes the nature of the wind and the resulting accretion flow. At high ionization luminosity ($L_\mathrm{X} = 10^{38}\;\mathrm{erg\ s}^{-1}$), the wind is strongly suppressed and the flow approaches the no-wind limit.

\item At moderate ionization ($L_\mathrm{X} = 10^{37}\;\mathrm{erg\ s}^{-1}$), the system enters a distinct accretion regime. The wind becomes denser and dynamically important near the compact object, strongly perturbing the RLO stream. As a result, the accretion flow transitions from an RLO-dominated to a wind-dominated regime.

\item In this moderate-ionization regime, the RLO stream undergoes a dynamical bifurcation into low- and high-angular-momentum branches. This bifurcation suppresses the formation of an extended accretion disk despite a continuous mass supply from the donor.

\item The effect of ionization is therefore non-monotonic: moderate ionization produces a more disruptive flow than both weak and strong ionization.

\item Additional simulations with a different ionization threshold ($\xi_0 = 30\;\mathrm{erg\ cm\ s}^{-1}$) show that the qualitative behavior is robust, although the strength of the wind-stream interaction is reduced due to stronger suppression of wind acceleration.

\end{enumerate}

These results demonstrate that the donor wind plays an active dynamical role in RLO HMXBs, not only by supplying mass but also by modifying the angular-momentum budget and flow structure. The interplay between wind dynamics and X-ray ionization introduces a new mechanism that can regulate accretion in these systems.

Future work should include a self-consistent treatment of the coupling between the X-ray luminosity and the mass accretion rate by performing simulations over timescales longer than the viscous timescales.
It will also be important to connect the present results with observational diagnostics, such as emission-line variability and X-ray spectral features, through radiative-transfer calculations.

\begin{ack}
Numerical computations were carried out on the general-purpose PC cluster at the Center for Computational Astrophysics, National Astronomical Observatory of Japan.
The authors used ChatGPT (OpenAI) during the preparation of this
manuscript to improve English wording, clarity, and organization.  The
scientific content, analytical derivations, numerical coefficients,
figures, interpretations, and conclusions were checked and finalized by
the authors, who take full responsibility for the content of the paper.
\end{ack}




\appendix

\section{Prescription for radiative acceleration}
\label{sec:radiative_acceleration}

To model the acceleration of stellar winds in our SPH simulations, we introduce an effective radiative force that reproduces a prescribed beta-type velocity law if no photoionizing source is present. Rather than solving the full line-driving problem, we adopt a simplified prescription in which the radiative acceleration is implemented as a modification of the gravitational force.

The equation of motion for a spherically symmetric flow is written as
\begin{equation}
v \frac{dv}{dr} = -\frac{GM}{r^2} + g_\mathrm{rad},
\label{eq:eom}
\end{equation}
where $v$ is the radial velocity, $r$ is the distance from the stellar center, $M$ is the stellar mass, and $g_\mathrm{rad}$ is the radiative acceleration.

In line-driven winds, the velocity distribution is often approximated by a beta velocity law \citep{LamersCassinelli1999},
\begin{equation}
v(r) = v_\infty \left(1 - \frac{R}{r}\right)^\beta,
\label{eq:beta_law}
\end{equation}
where $R$ is the stellar radius, $v_\infty$ is the terminal velocity, and $\beta$ is a parameter controlling the acceleration profile.
Substituting equation~(\ref{eq:beta_law}) into equation~(\ref{eq:eom}), we have
\begin{equation}
g_\mathrm{rad}(r) = \frac{GM}{r^2} +
                    \frac{\beta v_\infty^2 R}{r^2}
                    \left(1-\frac{R}{r}\right)^{2\beta-1}
\label{eq:g_rad_explicit}
\end{equation}
or
\begin{equation}
g_\mathrm{rad}(r) = \frac{GM}{r^2} \Gamma(r),
\label{eq:g_rad_gamma}
\end{equation}
where
\begin{equation}
\Gamma(r) = 1 + \frac{\beta v_\infty^2 R}{GM} \left(1 - \frac{R}{r}\right)^{2\beta - 1}.
\label{eq:beta_Gamma}
\end{equation}

In ionization cases, this effective radiative acceleration is multiplied by the reduction factor $f_\mathrm{ion}$ defined in equation~(\ref{eq:fion}).

This approach allows us to mimic the acceleration of line-driven winds without explicitly solving the line-force problem. Although the method does not include detailed microphysics such as ionization-dependent opacity variations, it provides a simple and computationally efficient prescription for reproducing a prescribed wind acceleration profile in SPH simulations.

\section{Effect of the threshold ionization parameter $\bm{\xi_0}$}
\label{sec:xi0_30_result}

\begin{figure}
\begin{center}
\includegraphics[width=0.25\textwidth]{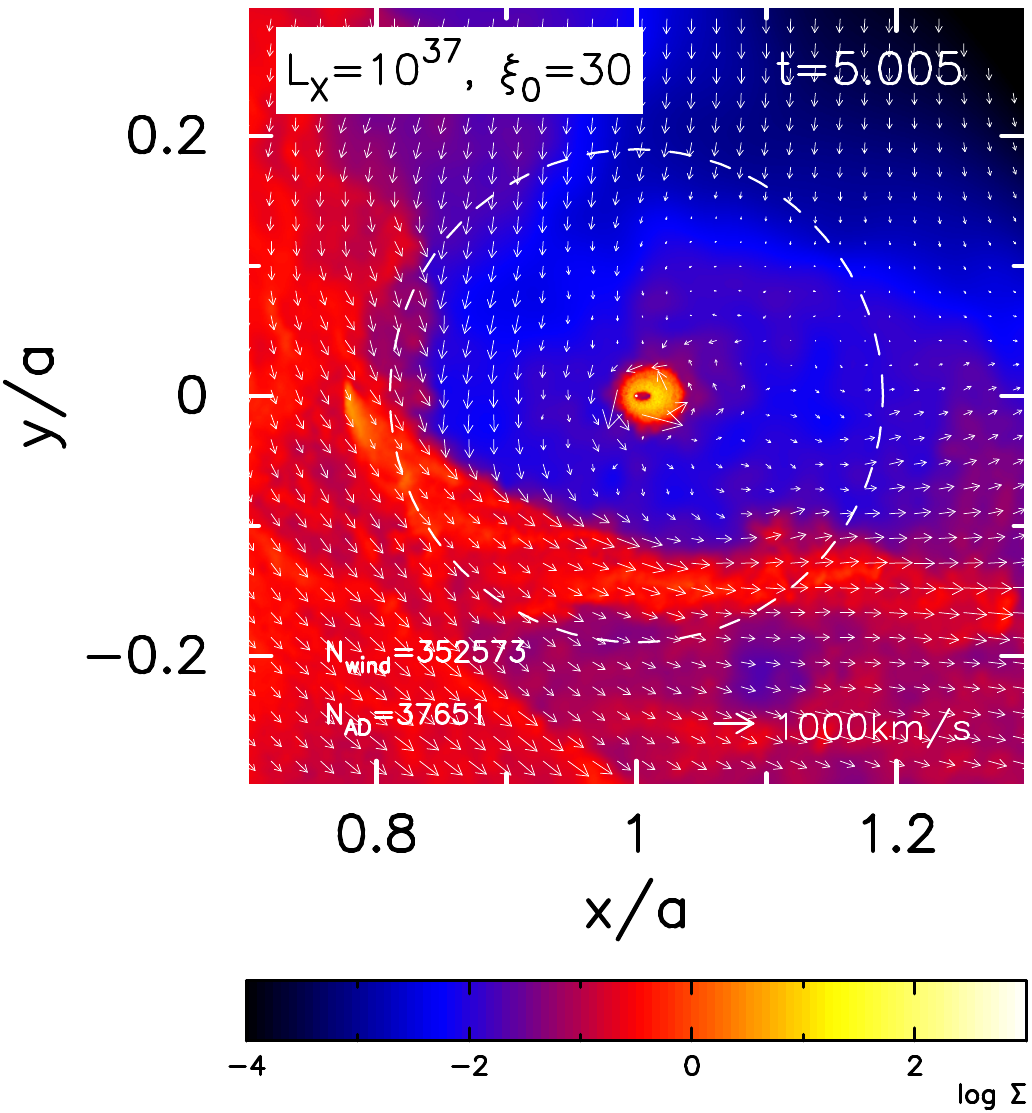}
\end{center}
\caption{
Surface density distribution and velocity field at the end of the moderate ionization ($L_\mathrm{X}=10^{37}\;\mathrm{erg\ s}^{-1}$) simulation with $\xi_0=30\;\mathrm{erg\ cm\ s}^{-1}$. The format of the figure is the same as that of Figure~\ref{fig:sigma+v}.
{Alt text: Surface-density map and velocity field for the moderate-ionization model with a lower ionization threshold at the end of the simulation. The flow shows a disturbed RLO stream and compact central structure, but the disk is larger than in the fiducial moderate-ionization model.}}
\label{fig:sigma+v_xi0_30}
\end{figure}

\begin{figure}
\begin{center}
\includegraphics[width=0.385\textwidth]{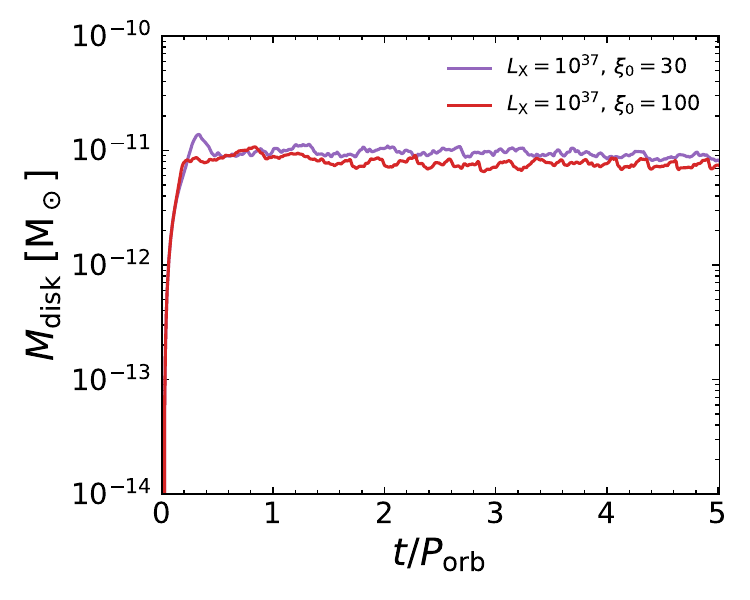}\hspace*{6mm} \\
\includegraphics[width=0.375\textwidth]{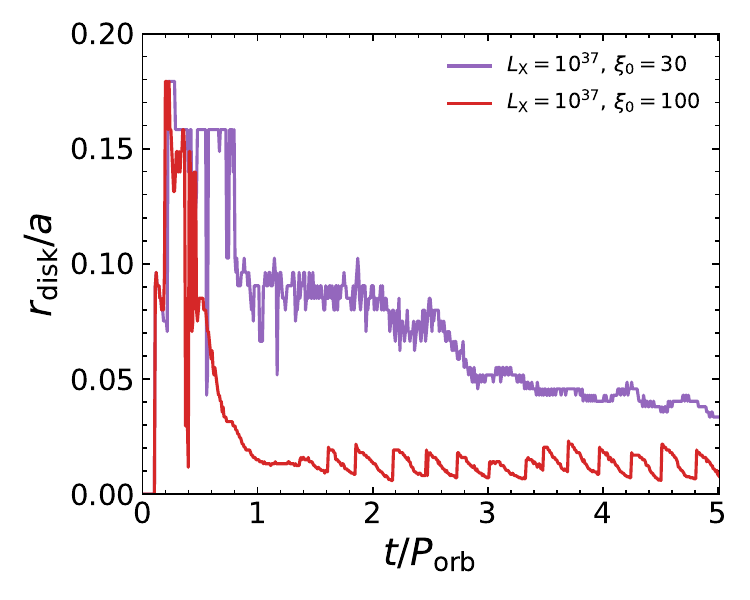}\hspace*{4mm} \\
\includegraphics[width=0.35\textwidth]{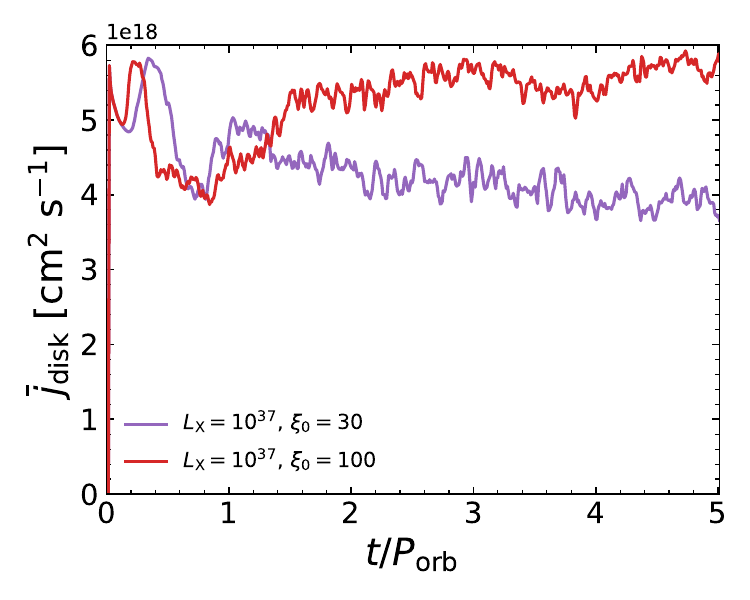}
\end{center}
\caption{
Time evolution of the disk mass (top), the disk radius (middle), and the mean specific angular momentum (bottom) of the disk for the moderate ionization ($L_\mathrm{X} = 10^{37}\;\mathrm{erg\ s}^{-1}$) models with $\xi_0=100\;\mathrm{erg\ cm\ s}^{-1}$ (red) and $\xi_0=30\;\mathrm{erg\ cm\ s}^{-1}$ (purple).
{Alt text: Time evolution of disk mass, disk radius, and mean specific angular momentum for two moderate-ionization models with different ionization thresholds. The low-threshold model has a disk mass similar to the fiducial model, but it maintains a larger disk radius and a lower mean disk specific angular momentum at late times.}}
\label{fig:evol_xi0_100_vs_30}
\end{figure}

\begin{figure}
\begin{center}
\includegraphics[width=0.45\textwidth]{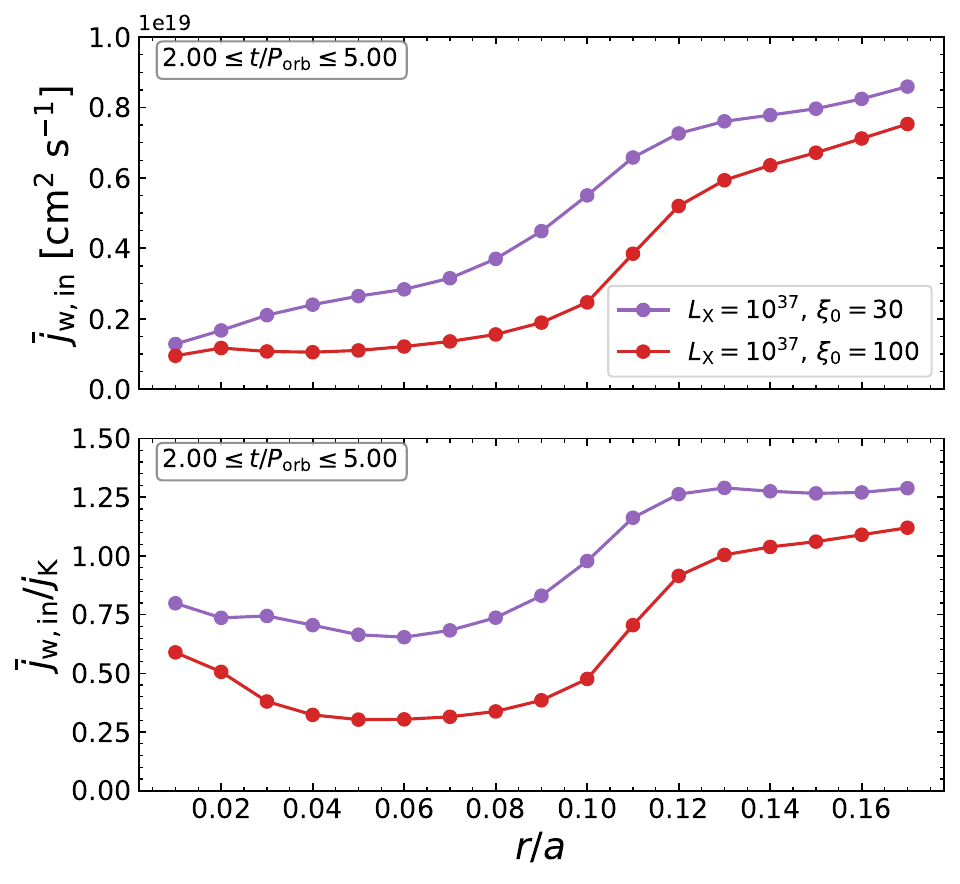}
\end{center}
\caption{
Radial profiles of the specific angular momentum carried by the inward wind around the compact object (upper panel) and its ratio to the local Keplerian value (lower panel) for the moderate-ionization ($L_\mathrm{X} = 10^{37}\;\mathrm{erg\ s}^{-1}$) models with $\xi_0=100\;\mathrm{erg\ cm\ s}^{-1}$ (red) and $\xi_0=30\;\mathrm{erg\ cm\ s}^{-1}$ (purple).
{Alt text: Radial profiles of the specific angular momentum carried by the inward wind for two moderate-ionization models with different ionization thresholds. The low-threshold model has a larger prograde inward-wind specific angular momentum than the fiducial model over most radii, indicating that the inward wind is more strongly biased toward the prograde direction.}}
\label{fig:windtorq_xi0_100_vs_30}
\end{figure}

To examine the sensitivity of our results to the ionization threshold,
we performed an additional simulation with $\xi_0 = 30\;\mathrm{erg\ cm\ s}^{-1}$ for the $L_\mathrm{X} = 10^{37}\,\mathrm{erg\ s^{-1}}$ model. In this appendix, we compare the resulting flow structure, disk evolution, and mean specific angular momentum distribution with those of the fiducial $\xi_0 = 100\;\mathrm{erg\ cm\ s}^{-1}$ model.
Figures~\ref{fig:sigma+v_xi0_30}, \ref{fig:evol_xi0_100_vs_30}, and \ref{fig:windtorq_xi0_100_vs_30}
show the global flow structure, the disk evolution, and the radial profiles of mean specific angular momentum carried by the inward wind, respectively.

In general, results for $\xi_0 = 30\;\mathrm{erg\ cm\ s}^{-1}$ are qualitatively similar to those for $\xi_0 = 100\;\mathrm{erg\ cm\ s}^{-1}$. However, because the smaller ionization threshold suppresses the wind acceleration more efficiently, less wind material reaches the vicinity of the neutron star. As a result, the wind-stream interaction becomes weaker than in the $\xi_0 = 100\;\mathrm{erg\ cm\ s}^{-1}$ case.

Compared with the $\xi_0 = 100\;\mathrm{erg\ cm\ s}^{-1}$ model, the accretion disk in the $\xi_0 = 30\;\mathrm{erg\ cm\ s}^{-1}$ model is larger and more stable, the wind is more dominated by the prograde component, and the bifurcation of the specific angular momentum distribution becomes less pronounced.

These results suggest that the disturbed-flow regime found at moderate ionization is robust against moderate changes in the ionization threshold, although the quantitative strength of the wind-stream interaction depends sensitively on the ionization prescription.

\end{document}